\documentclass[aps,nofootinbib,superscriptaddress, showpacs,preprintnumbers,  nofootinbibt,twocolumn]{revtex4-1}
\usepackage{epsfig}
\usepackage{multirow}
\usepackage{eurosym}
\usepackage{dcolumn}
\usepackage{bm}
\usepackage{enumerate}
\usepackage{float}
\usepackage{epstopdf}
\usepackage{amsmath}
\usepackage{bm}
\usepackage{amsfonts}
\usepackage{amssymb}
\usepackage{graphicx}
\usepackage{alphalph,mathtools}
\usepackage{etoolbox}
\usepackage{color}
\usepackage{booktabs}
\usepackage{footnote}
\usepackage{makecell,tabularx}

\def\be{\begin{equation}}
\def\ee{\end{equation}}
\def\bea{\begin{eqnarray}}
\def\eea{\end{eqnarray}}

\newcommand{\cH}{\ensuremath{\mathcal{H}}}

\begin{document}

\title{Cosmological tests of the osculating Barthel-Kropina dark energy model}
\author{Amine Bouali}
\email{a1.bouali@ump.ac.ma}
\affiliation{Laboratory of Physics of Matter and Radiation, Mohammed I University, BP 717, Oujda, Morocco,}
\author{Himanshu Chaudhary}
\email{himanshuch1729@gmail.com}
\affiliation{Department of Applied Mathematics, Delhi Technological University, Delhi-110042, India,}
\author{Rattanasak Hama}
\email{rattanasak.h@psu.ac.th}
\affiliation{Faculty of Science and Industrial Technology, Prince of Songkla University,
Surat Thani Campus, Surat Thani, 84000, Thailand,}
\author{Tiberiu Harko}
\email{tiberiu.harko@aira.astro.ro}
\affiliation{Department of Physics, Babes-Bolyai University, Kogalniceanu Street,
	Cluj-Napoca, 400084, Romania,}
\affiliation{Department of Theoretical Physics, National Institute of Physics
and Nuclear Engineering (IFIN-HH), Bucharest, 077125 Romania,}
\affiliation{Astronomical Observatory, 19 Ciresilor Street,
	Cluj-Napoca 400487, Romania,}
\author{Sorin V. Sabau}
\email{sorin@tokai.ac.jp}
\affiliation{School of Biological Sciences, Department of Biology, Tokai University, Sapporo 005-8600, Japan}
\affiliation{Graduate School of Science and Technology, Physical and Mathematical Sciences, \\
Tokai University, Sapporo 005-8600, Japan,}
\author{Marco San Mart\'in }
\email{mlsanmartin@uc.cl}
\affiliation{Faculty of Engineering, Pontifical Catholic University of Valparaiso, Chile}

\begin{abstract}
We further investigate the dark energy model based on the Finsler geometry inspired osculating Barthel-Kropina cosmology. The Barthel-Kropina cosmological approach is based on the introduction of a Barthel connection in an osculating Finsler geometry, with the connection having the property that it is the Levi-Civita connection of a Riemannian metric. From the generalized Friedmann equations of the Barthel-Kropina model, obtained by assuming that the background Riemannian metric is of the Friedmann-Lemaitre-Robertson-Walker type, an effective geometric dark energy component can be generated, with the effective, geometric type pressure, satisfying a linear barotropic type equation of state. The cosmological tests, and comparisons with observational data of this dark energy model are considered in detail. To constrain the Barthel-Kropina model parameters, and the parameter of the equation of state, we use 57 Hubble data points, and the Pantheon Supernovae Type Ia data sample. The st statistical analysis is performed by using Markov Chain Monte Carlo (MCMC) simulations. A detailed comparison with the standard $\Lambda$CDM model is also performed, with the  Akaike information criterion (AIC), and the Bayesian information criterion (BIC) used as the two model selection tools. The statefinder diagnostics consisting of jerk and snap parameters, and the $Om(z)$ diagnostics are also considered for the comparative study of the Barthel-Kropina and $\Lambda$CDM cosmologies. Our results indicate that the Barthel-Kropina dark energy model gives a good description of the observational data, and thus it can be considered a viable alternative of the $\Lambda$CDM model.
\end{abstract}

\date{\today }
\maketitle
\tableofcontents

%\preprint{APS/123-QED}

\section{Introduction}

Finsler geometry \cite{F1b} is an interesting and important extension of Riemannian geometry \cite{r1}.  The relationship between the two geometries was best described by
Chern, who described it as  "Finsler Geometry is just Riemannian geometry without the quadratic restriction" \cite{Chern}. Indeed, in his  Habilitationsvortrag \cite{r1} Riemann has already introduced a metric structure in a general space based on the invariant distance element $ds=F\left(x^1,x^2,...,x^n;dx^1,dx^2,..., dx^n\right)=F(x,y)$, where for $y\neq 0$, $F$ is a positive definite function defined on the tangent bundle $TM$. Moreover, $F$ is assumed to be homogeneous of degree one in $y$. Riemannian geometry is a special case of the general metric, with $F^2=g_{ij}(x)dx^idx^j$. Hence, in a proper sense, Finsler geometry is not a generalization of the Riemannian geometry, but it is the Riemannian geometry without the quadratic restriction \cite{Chern}. However, it is customary in both mathematical and physical literature to make a clear distinction between Riemann and Finsler geometries, and in the present work we will adopt the standard terminology. In a physically more intuitive way we may consider Finsler geometry as a geometry in which the metric tensor is a function of both coordinates defined on the base manifold, and of the tangent vectors, $g_{ij}=g_{ij}(x,y)$, or, as a geometry in which the metric tensor is a function of both coordinates and velocities. For in depth presentations and discussions of Finsler geometry see \cite{F2b,F3b,F4b,F5b,F6b}.

Despite its systematic and rigorous nature, and its many attractive features, the physical applications of the Finsler geometry were shadowed by the immense successes of the theory of general relativity, which is essentially based on Riemannian geometry \cite{Ein0,Ein1, Hil}. The advent of general relativity also led to important developments in mathematics, like Weyl geometry \cite{Weyl}, geometries with torsion \cite{Cartan1,Cartan2}, or geometries with absolute parallelism \cite{Weitz}. All these geometries have found important applications in physics, and opened new perspectives in the understanding of the gravitational interaction. On the other hand, the applications of the Finsler geometry to physics did appear relatively late. One important step in this direction was taken in the work by Randers \cite{Rand}, initially still formulated in a higher dimensional Riemannian context, with the goal of obtaining a unified theory of gravity and electromagnetism. However, Randers geometry is a typical example of a Finsler geometry, with $F(x,y)=\left(a_{ij}(x)dx^idx^j\right)^{1/2}+A_k(x)y^k$, where $A_k(x)$ is an arbitrary vector field. Recently, Randers geometry was extensively applied in the study of various gravitational phenomena in \cite{R1a, R2a,R3a,R4a,R5a,R6a,R7a,R8a}. Finsler geometry has also important applications in the geometric description of quantum mechanics \cite{Tave1,Tave2,Tave3,Tave4}. General relativistic kinetic gas theory was investigated by using methods from Finsler geometry in \cite{Voicu2}.

The first attempts at formulating a Finslerian theory of gravitation belonged to Horv\'{a}th \cite{Hor1b}, and Horv\'{a}th and Mo\'{o}r \cite{Hor2b}.   Early Finslerian type gravitational theories were also formulated in \cite{Hor3b} and \cite{Hor4b}, respectively. The set of the Finslerian type gravitational field equations proposed in these works are given by
\be\label{h1}
R_{\mu \nu}-\frac{1}{2}g_{\mu \nu}R+\lambda g_{\mu \nu}=\chi T_{\mu \nu},
\ee
\be\label{h1a}
K_{\mu \nu}-\frac{1}{2}g_{\mu \nu}K+\lambda g_{\mu \nu}=\chi T_{\mu \nu},
\ee
and
\be\label{h3}
S_{\mu \nu}-\frac{1}{2}g_{\mu \nu}S-\lambda^{(i)} g_{\mu \nu}=-\chi ^{(i)}T_{\mu \nu}^{i},
\ee
respectively, where $\chi$ is the gravitational constant, $\lambda$ is the cosmological constant, while $\lambda ^{(i)}$ and $\chi ^{(i)}$ denote the internal cosmological and gravitational constants, respectively. The geometrical quantities  $R_{\mu \nu}$, $R$, $K_{\mu \nu}$, $K$ are the contracted third curvatures, while $S_{\mu \nu}$ and $S$ are the v-Ricci curvature tensors, and the v-scalar curvature, respectively. Moreover, $T_{\mu \nu}$ denotes the ordinary matter energy-momentum tensor, while $T_{\mu \nu}^{i}$ is the internal energy-momentum tensor.

An interesting Finslerian approach to gravity was introduced in \cite{As0}, with the main focus on the Finslerian interpretation of the particle motion in a gravitational field. For detailed presentations of the Finslerian type extensions of general relativity see \cite{As1}. The Schwarzschild type Finslerian metrics were considered in \cite{As2,As3}.

The vector bundle point of view was adopted to propose a system of Einstein type gravitational field equations in \cite{Miron}. The basic idea of this approach is to consider the field  $y$ as a fibre at the point $x$ of the base $x$ manifold. For this vector bundle the total space is constructed  as a unification of the $x$ and $y$ fields \cite{Ikeda}. The adapted frame is defined for these unified fields as
\bea\label{eq3}
X_A&=&\left(\frac{\delta}{\delta x^{\lambda}}=\frac{\partial}{\partial x^{\lambda}}-N_{\lambda}^i\frac{\partial}{\partial y^i}, \frac{\partial}{\partial y^i}\right),\nonumber\\
X^A&=&\left(dx^{\kappa},\delta y^i=dy^i+N_{\lambda}^idx^{\lambda}\right),
\eea
where $X_A$ is the adapted basis of the tangent
space $T_xM$, while $X^A$ is the adapted cobasis in the cotangent
space $T_x^*M$. In Eq.~(\ref{eq3}) the indices $A,B$ take the values  $A,B=\left(\kappa,i\right)\in \left\{0,1,2,3,...,7\right\}$, while $\lambda, \kappa= 0,1,2,3$. Moreover, $N_{\lambda}^i$ denotes the nonlinear connection. The frame (\ref{eq3}) is adapted to the metric $G=\left(G_{AB}\right)$, given by
$G=g_{\lambda \kappa}(x,y)dx^{\kappa}dx^{\lambda}+g_{ij}(x,y)\delta y^i\delta y^j$.
On the total space the gravitational field equations are postulated to have the standard form, $ \mathcal{R}_{AB}-(1/2)\mathcal{R}G_{AB}=\tau _{AB}$,
 where $\tau _{AB}$ is the matter energy-momentum tensor, and they can be decomposed to take the form \cite{Miron}
\be
R_{\lambda \nu}-\frac{1}{2}(R+S)g_{\lambda \nu}=\tau _{\lambda \nu},
\overset{1}{P}_{i \lambda}=\tau _{i\lambda},  \overset{2}{P}_{ \lambda i}=-\tau _{\lambda i},
\ee
\be
 S_{i j}-\frac{1}{2}(R+S)g_{i j}=\tau _{ij}.
\ee

An alternative approach to the Finslerian geometric type gravity was proposed in \cite{Rutz}, by assuming that the Einstein vacuum equations are given by $H=H_i^i=0$, with $H_k^i$ constructed from the first and second derivatives of the quantity $G^l=\gamma _{jk}^l\dot{x}^j\dot{x}^k/2$. For a Riemannian metric, the gravitational field equation reduce to the general relativistic Einstein gravitational field equations. Finslerian type solutions of the field equations can also be obtained.

An interesting and important class of Finsler space are the Berwald-Finsler spaces. In these geometries a particular set of gravitational field equations
was introduced and discussed in \cite{Lixin}. To obtain the gravitational field equations in the Berwald-Finsler  space the Bianchi identities satisfied by the Chern curvature have been used. The geometric part of the gravitational field equation is nonsymmetric in general, indicating that the principle of the local Lorentz invariance is not satisfied.

Finsler type gravitational field equations have been obtained from a Finsler-Lagrange function $L$ in \cite{Voicu1}, with the use of a variational principle, The field equations are given by
\be
2R - \frac{L}{3}g^{Lij}R_{\cdot i \cdot j} + \frac{2L}{3}g^{Lij}\left[ (\nabla P_{i})_{\cdot j} + P_{i|j} - P_{i}P_{j}\right]= 0,
\ee
where $g^L_{ij}=\frac{1}{2}\frac{\partial ^{2}L}{\partial \dot{x}^{i}\partial \dot{x}^{j}}=\frac{1}{2} L_{\cdot i\cdot j}$, $R_{.i.j}$ is the geodesic deviation operator, $R$ is its trace, and $P$ is the Landsberg tensor. The action from which the above field equations can be derived is given by $S[L] = \int_{\Sigma\subset TM} \mathrm{vol}(\Sigma) R_{|\Sigma}$, where $\Sigma = \{(x,\dot x)\in TM|F(x,\dot x) = 1\}$ is the unit tangent bundle, and $\mathrm{vol}(\Sigma)$ is the volume form on $\Sigma$, constructed with the use of the Finsler metric.

In its Riemannian formulation the theory of General Relativity was extremely successful in explaining the gravitational phenomenology at the level of the Solar System, where it passes all high precision observational tests, including the light deflection, the perihelion advance of Mercury,  the Shapiro time delay, the Nordtvedt effect in lunar motion, and frame-dragging, respectively \cite{Will}. An important confirmation of the predictions of general relativity is represented by the experimental detection of the gravitational waves \cite{Grav1}, which opened a deep view on the Universe, also leading, for example, to a new perspective on the mass distribution of the neutron stars \cite{Grav2}. However, when extended to gravitational systems far bigger than the Solar System, namely, at the galactic and cosmological scales, General Relativity is facing a number of very serious difficulties, whose solutions may require a fundamental change in our view of the gravitational interaction.

The precise measurements of the temperature fluctuations of the Cosmic Microwave Background Radiation (CMBR) by the Planck satellite \cite{1g,1h}, together with the astrophysical  observations of the distant supernovae, extending up to a redshift of $z\approx 2$ \cite{Ri98,Ri98-1,Ri98-2,Ri98-3, Ri98-4,Hi,Ri98-5, Ri98-6} have firmly indicated that the present-day Universe is in a phase of accelerating expansion. Moreover,  the amazing result that its matter content consists of only 5\% baryonic matter has also been confirmed. Hence, these observations strongly point out that 95\% of the matter-energy content of the Universe consists of two main (and mysterious) components, dark energy and dark matter, respectively.

An explanation of these cosmological observations can be obtained if one  reintroduces in the Einstein  gravitational field equations of the cosmological constant $\Lambda$, first introduced by Einstein in 1917 \cite{Einb}. For discussions about the history of the cosmological constant, and its possible interpretations see \cite{W1b,W2b,W3b}. The corresponding cosmological model, obtained by also adding a cold dark matter component in the Einstein field equations is called the $\Lambda$CDM model, and it has become one of the main theoretical tool for the understanding of the cosmic dynamics.

The $\Lambda$CDM paradigm can give excellent fits to the observational data. However, due to the lack of a convincing theoretical basis, and of the many problems raised by the cosmological constant itself, the physical basis of the $\Lambda$CDM model is (at least) uncertain. Therefore, to obtain a mathematically, physically and observationally consistent description of the Universe, three major, and distinct, theoretical approaches have been proposed, called the dark components model, the dark gravity model, and the geometry-matter coupling model, respectively \cite{LoHa}. The dark components model \cite{Rev1,Rev2,Rev3,Rev4, Rev5} postulates that the basic constituents of Universe are dark energy, and dark matter, respectively, of (yet) unknown physical origin. There are many proposals for the nature of these dark constituents. A simple dark energy model can be obtained in the framework of the quintessence models  \cite{quint1b,quint2b,quint3b,quint4b, quint5b},  in which the dynamics of the Universe is determined by a single self-interacting scalar field $\phi$, in the presence of a potential $V(\phi)$.  For the quintessence models the gravitational action is given by
\be
S=\int{\left[\frac{M_{p}^2}{2}R-\partial_{\mu} \phi\partial ^\mu-V(\phi)\right]\sqrt{-g}d^4x},
\ee
where $R $ is the Ricci scalar, and $M_p$ denotes the Planck mass.

In the dark gravity approach it is proposed  that to explain the gravitational interaction that changes on astrophysical (galactic) and cosmological scales, one must go beyond the Riemannian geometry of general relativity,  and that more general geometries must be used for a theoretical description of gravity. In this direction theories in the presence of torsion \cite{Tor1,Tor2,Tor3,Tor4}, of nonmetricity  \cite{Nest1,Nest2,Nest3, Ghil1,Ghil2,Ghil3}, or in the Weitzenb\"{o}ck geometry \cite{We1,We2} have been intensively investigated. The third theoretical approach to the gravitational phenomenology assumes that ordinary matter may play a dominant role in the cosmological dynamics due to its coupling with geometry, via a curvature - matter coupling  that could explain the observed gravitational phenomenology \cite{Co1,Co2,Co3,Co4,Co5}. For reviews of modified gravity models see \cite{R1,R2,R3,R4,R5}. For a detailed analysis of theories with geometry-matter coupling see \cite{e8}.

 In the dark gravity approach, which goes beyond the mathematical formalism of the Riemann spaces,  Finsler type cosmological models represent an interesting alternative to the standard $\Lambda$CDM model, as geometric explanations, or replacements, of dark energy, and perhaps even of dark matter. Many studies have been devoted to the applications of the Finsler geometry in cosmology, with the goal of understanding from a new point of view the dynamical evolution of the cosmic structures  \cite{Fc1,Fc2,Fc3,Fc4,Fc5,Fc6,Fc7,Fc8,Fc9,Fc10,Fc11,Fc12,Fc13,Fc14,Fc15,Fc16,Fc17,Fc18,Fc19,Fc19a, Fc20,Fc21,Fc22,Fc23, Fc24, Fc24a, Fc25,Fc26}.

 In particular, in \cite{Fc26}, the cosmological implications of the Kropina geometry have been investigated in detail, by using the mathematical formalism of the osculating Finsler spaces, in which the internal variable is a function of the base manifold coordinates only. Moreover, in order to describe gravitational phenomena, the Barthel connection was adopted, which has the remarkable property that it is the Levi-Civita connection of a Riemannian metric. To describe the gravitational phenomena it was assumed that in the Barthel-Kropina geometry the Ricci type curvatures are related to the matter energy-momentum tensor by the standard Einstein equations. The generalized Friedmann equations in the Barthel-Kropina geometry have been derived by considering that the background Riemannian metric is of Friedmann-Lemaitre-Robertson-Walker (FLRW) type. The model admits a de Sitter type solution, and an effective fluid type dark energy component, described by an effective energy density and thermodynamic pressure, related by a linear equation of state,  can also be obtained.  A preliminary comparison of the dark energy model with the observational data and with the standard $\Lambda$CDM model was also performed, and it was found that the Barthel-Kropina-FLRW type models give a satisfactory description of the observations.

 It is the goal of the present paper to perform a detailed comparison between the theoretical predictions of the Barthel-Kropina dark energy model, and the cosmological observations. The Barthel-Kropina dark energy model contains three free parameters, the coefficient of the equation of state $\omega$, and the present day values of the coefficient of the one form $\beta$, and of its derivative, respectively. To find the values, and the constraints, on these parameters, we used the statistical MCMC approach,  with the Bayesian technique. Moreover, we fit the theoretical predictions with two different observational samples, containing Hubble data, and the Pantheon data. Hence, we can obtain in this way the best fit values of the model parameters. With the fitted values of the parameters we perform a cosmographic (statefinder) analysis of the model, by investigating the behaviors of the deceleration parameter, and of the jerk and snap parameters, respectively. The $Om(z)$ diagnostic is also considered. In all cases, by using the fitted values of the model parameters, we compare the Barthel-Kropina model with the standard $\Lambda$CDM model.

 The present paper is organized as follows. In Section~\ref{sect1} we briefly review the basic Finsler geometric concepts used in the construction of the gravitational model, including the definitions of the Barthel connection, and of the osculating Finsler geometry. The basic principles of the Barthel-Kropina cosmology are introduced in Section~\ref{sect2}, where the generalized  Friedmann equations are also written down. We present the Barthel-Kropina dark energy model, and discuss its properties in Section~\ref{sect3}. A detailed comparison of the theoretical predictions of the model and the observational data is performed in Section~\ref{sect4}. We discuss and conclude our results in Section~\ref{sect5}.

 \section{Quick introduction to Finsler geometry, $\left(\alpha, \beta\right)$ metrics, and the Barthel connection}\label{sect1}

In the present Section we will briefly review the basic concepts of the Finsler geometry to be used in the cosmological applications. Specifically, we focus on the concept of Finsler and $(\alpha,\beta)$-metrics. In our approach the Kropina metric plays a central role, and therefore we will consider it in some detail. Our investigations are essentially based on the concept of osculating Finsler spaces, and of the Barthel connection, respectively, which makes necessary a brief presentation of the basics of the Barthel-Kropina geometry. For in depth presentations of the Finsler geometry, and of some of its applications, see \cite{F2b,F3b,F4b,F5b,F6b}.

 Finsler geometry has many applications in classical Newtonian physics, especially in the description of the dissipative effects. In classical mechanics the
equations of motion of a system of particles in the presence of external forces $F_i$, $i=1,2,...,n$, which cannot be derived from a potential,  can be obtained from a regular Lagrangian $%
L$, defined on an $n$-dimensional differentiable manifold $M$, by using the Euler-Lagrange equations, given by
\begin{equation}
\frac{d}{dt}\frac{\partial L}{\partial y^{i}}-\frac{\partial L}{\partial
x^{i}}=F_{i},i\in \left\{1,2,...,I\right\},  \label{EL}
\end{equation}
The Euler-Lagrange
equations  (\ref{EL}) are equivalent to a system of second-order differential equations,
\begin{equation}\label{EM}
\frac{d^{2}x^{i}}{dt^{2}}+2G^{i}\left( x,y,t\right) =0,i\in \left\{
1,2,...,n\right\} .
\end{equation}
From a mathematical point of view, Eqs.~(\ref{EM}) describe geodesic motion in a Finsler space.

\subsection{Finsler geometry, and particular Finsler spaces}

In the present day approach to the basic laws describing natural phenomena a basic assumption is that time and space form together a single structure, called the space-time. Mathematically, the space-time is described  as a four dimensional differentiable manifold $M$, on which a pseudo-Riemannian metric tensor $g_{I
J} $,  $I,J,K...=0,1,2,3$, can be defined. According to the chronological hypothesis, the space-time distance (interval) between two
events $x^{I}$ and $x^{I} + dx^{I}$  is obtained according to the prescription $ds=\left(g_{IJ}dx^{I}dx^{J}\right)^{1/2}$  \cite%
{Tav,Tav1}. In Riemannian geometry, the metric tensor $g_{IJ}$ is a function of the coordinates $x^I$ of the space-time manifold only, so that $g_{IJ}=g_{IJ}(x)$. But more general geometries than the Riemannian one can also be constructed. One of the important metrical generalizations of the Riemann geometry is the Finsler geometry \cite{F1b,F2b,F3b,F4b,F5b, F6b}.

 Finsler spaces are a class of metric spaces, in which the distance $ds$ between two neighbouring points $x=(x^{I})$ and $x+dx=(x^{I} + dx^{I})$ is obtained according to the relation,
\begin{equation}\label{dsF}
d\hat{s}=F\left(x,dx\right).
\end{equation}
The Finsler metric function $F$, introduced in  the above definition, is a  positively homogeneous of degree
one function in $dx$, with the basic property
\begin{equation}
F\left(x,\lambda dx\right)=\lambda F\left(x,dx\right), {\rm for} \lambda >0.
\end{equation}

The Finsler metric function $F$ is usually expressed by using the
canonical coordinates $(x,y)=(x^I,y^I)$ of the tangent bundle TM, where $y=y^I\dfrac{\partial}{%
\partial x^I}$, is a tangent vector at $x$. By using the canonical coordinates, the Finsler metric tensor $%
\hat{g}_{I J}$ is defined according to
\begin{equation}\label{Hessian mat}
\hat{g}_{I J}\left(x,y\right)=\frac{1}{2}\frac{\partial ^2F^2\left(x,y\right)}{%
\partial y^{I}\partial y^{J}},
\end{equation}
on the tangent bundle TM=TM$\backslash$0.
Hence, we can write Eq.~(\ref{dsF}) as $d\hat{s}^2=\hat{g}_{I J}\left(x,y\right)y^{I}y^{J}$.
Riemann spaces can be considered as particular cases of Finsler spaces, obtained when $\hat{g}_{IJ}\left(x,y\right)=g_{IJ
}\left(x\right)$, $y^{I}=dx^{I}$, leading to $ds^2=g_{IJ}(x)dx^Idx^J$, respectively. For a discussion on the relation between Riemann and Finsler geometries see \cite{Chern, F5b}.

One can also introduce an important additional geometric quantity, the Cartan tensor $\hat{C}(x,y)$, which is defined as
\begin{equation}
\hat{C}_{IJK}=\frac{1}{2}\frac{\partial \hat{g}_{IJ}\left( x,y\right) }{\partial
{y}^{K}}.
\end{equation}

The Cartan tensor gives an estimation of the deviation of a Finsler geometry from a Riemannian one.

\subsubsection{Particular Fisnler geometries - Randers, Kropina and general $(\alpha , \beta )$ metrics}

There are a large number of special Finsler geometries, obtained by specifying the functional form of the metric tensor $\hat{g}_{I J}\left(x,y\right)$. One of the first considered Finsler geometries, which has many applications in physics, are the Randers spaces \cite{Rand}, representing a special type of Finsler structures, with the Finsler metric function defined as
\begin{equation}
F=\left[ g_{I J }(x)dx^{I }dx^{J }\right] ^{1/2}+A_{I }(x)dx^{I }=\alpha +\beta,
\end{equation}%
where $g_{I J }(x)$ is the metric tensor of a {\it Riemann space},
and $A_{I}(x)dx^{I }$ is a linear $1$-form, defined on the tangent bundle $TM$.

Another important class of Finsler geometries are the Kropina spaces \cite{Krop,Krop1}, which are special Finsler spaces with metrics given by
\begin{equation}
F\left( x,y\right) =\frac{g_{IJ }(x)y^{I }y^{J }}{A_{I }(x)y^{I }}.
\end{equation}

A generalization of the above geometries was done by  Matsumoto \cite{Mat,Mat1}, by introducing  {\it the concept
of the} $(\alpha ,\beta )$ {\it metrics}.  An $(\alpha ,\beta )$ metric is obtained when the Finsler metric function $F$ {\it is a positively
homogeneous function $F(\alpha ,\beta )$ of first degree in two variables} $%
\alpha \left( x,y\right) =\left[ g_{I J }(x)dx^{I }dx^{J }\right] ^{1/2}$
and $\beta \left( x,y\right) =A_{I }(x)y^{I }$, respectively.

As for $\alpha $, we assume that it is a non-degenerate (regular), and positive-definite{\it  Riemannian metric}. Both the Randers and the Kropina metrics
belong to the class of the $(\alpha ,\beta )$ metrics. In the case of the Randers metric $F=\alpha +\beta $, while $F=\dfrac{\alpha^2}{\beta}$ for the Kropina metric. One can also define general $(\alpha ,\beta )$ metrics, with the Finsler metric function given by $F(\alpha ,\beta )=\alpha \phi \left( \beta /\alpha \right) =\alpha \phi \left(s\right)$, where $ s=\beta /\alpha $, and $\phi =\phi (s)$ is a $ C^{\infty }$ positive function, defined on an open interval $(-b_{o},b_{o})$ \cite{Fc26}.

For the fundamental metric tensor of the $(\alpha, \beta)$ metric  we obtain the expression
\bea
\hat{g}_{IJ}(x,y)&=&\frac{L_{\alpha }}{\alpha }h_{IJ}+\frac{L_{\alpha \alpha }%
}{\alpha ^{2}}y_{I}y_{J}+\frac{L_{\alpha \beta }}{\alpha }\left(
y_{I}A_{J}+y_{J}A_{I}\right) \nonumber\\
&&+L_{\beta \beta }A_{I}A_{J},
\eea
where we have denoted $L=F^{2}/2$, and
\be
h_{IJ}=\alpha \frac{\partial ^2\alpha (x,y)}{\partial y^I \partial y^J}=g_{IJ}-\frac{y_Iy_J}{\alpha ^2}.
\ee
As usual, the indices $\alpha $, $\beta $ of $L$ denote partial differentiation
with respect to $\alpha $ and $\beta$, respectively.

\subsection{From the Barthel connection to the $Y$-osculating Riemann spaces}

We now briefly present the two basic mathematical concepts on which the cosmological applications of the Kropina metrics, considered in the present study,  are based, namely, the Barthel connection, and the osculating Finsler spaces.

\subsubsection{The Barthel connection}

Let's consider now that $(M^n,F)$ is a Finsler space, defined on a base manifold $M^n$. On $M^n$ a vector field $Y(x)\neq 0$ is also defined.
We can define now a particular mathematical structure $(M^n,F(x,y),Y(x))$, representing a Finsler space %
$(M^n,F(x,y))$ {\it which has a tangent vector field $Y(x)$}. {\it For a vector $Y$ that does not vanish in any point on $M$, the Finslerian metric $\hat{g}(x,y)$ generates the $Y$-Riemann metric $%
\hat{g}_{Y}(x)=\hat{g}(x,Y)$.}

Point Finsler spaces represent an important class of Finsler spaces.   {\it A point Finsler space is an} $n$-{\it dimensional space, which is locally Minkowskian, and, in general, not locally Euclidean} \cite{Ba1,Ba2}.  A general Finsler geometry is {\it inhomogeneous and anisotropic}, while a Minkowski space is {\it flat, homogeneous, but still anisotropic}.  {\it The Finsler } $n${\it -space can be called a Barthel-Finsler space}, or a {\it point Finsler space}.

Let's assume that a point vector field $Y^I(x)$ and a Finsler metric tensor $\hat{g}(x,y)$ are given. Then, we define the absolute differential of the vector $Y$ as \cite{In1}
\be\label{DY}
DY^I=dY^I+Y^K b^I_{KH}(x, Y) dx^H,
\ee
where $b^I_{KH}(x, Y)$ denotes {\it the coefficients of the Barthel connection}. The coefficients $b^I_{KH}(x, Y)$ are obtained with the help of the generalized Christoffel symbols $\hat{\gamma}_{JIH}$, defined, as in Riemann geometry, according to the relation
\be
\hat{\gamma}_{IJH}:=\frac{1}{2}\left(\frac{\partial \hat{g}_{JI}}{\partial x^H}+\frac{\partial \hat{g}_{IH}}{\partial x^J}-\frac{\partial \hat{g}_{HJ}}{\partial x^I}\right).
\ee

To obtain the explicit expression of the Barthel connection $b^I_{KH}(x, Y)$ we write the expressions in the second term of Eq.~(\ref{DY}) as
\be
Y^Kb^I_{KH}=Y^K\left(\hat{\gamma}^I_{KH}-\hat{\gamma}^R_{KS}Y^S\hat{C}^I_{RH}\right),
\ee
which allows us to obtain the Barthel connection as \cite{In1}
\be
b^I_{KH}=\hat{\gamma}^I_{KH}-\hat{\gamma}^R_{KS}Y^S\hat{C}^I_{RH}.
\ee

The Barthel connection has several interesting properties. First of all,  {\it it depends on the vector field on which it acts}, a property that does not appear in Riemann geometry. Hence, the Barthel connection is very different, as compared to the connections in Riemann geometry. Generally, for anisotropic metrics, all geometric properties do depend on the direction. However,  {\it for the Barthel connection, the dependence is only on the direction of the vector field, and not on its magnitude}. Moreover, the Barthel connection {\it is the simplest connection that keeps the metric function unchanged by the parallel transport}. In the case of Finsler vector fields, which are functions of both $x$ and $y$, the Barthel connection permits a natural transition to the Cartan  geometry of the Finsler spaces. Hence, in the following, we consider {\it the connection of a point Finsler space as the Barthel connection}.

 The Barthel connections, unlike the usual Levi-Civita connection of a Riemannian metric, or, more general affine connections, do not live on the base manifold $M$, but on the total space of the tangent bundle \cite{F5b,F6b}. This important characteristic may lead to major differences between the geometrical theories of Finsler and Riemann manifolds.

\subsubsection{The $Y$-osculating Riemann geometry}

The concept of osculating Riemann spaces of Finsler geometries was developed by Nazim \cite{Naz}, and it was later studied in great detail in \cite{Varga}.  The osculating approach associates to a complex geometric structure, like, for example, a Finsler geometry, and a Finsler connection, a simpler mathematical format, like a Riemann metric, or an affine or a linear connection. In doing this one assumes that the simpler, osculating structure, approximates, in some sense, the most complicated one. Hence, by using the osculation approach, one can obtain mathematical results that allow the understanding of the properties of the mathematically more complicated geometries.

Let's consider now a local section $Y$ of $\pi_M:TM\to M$. It is a basic mathematical result that geometric objects defined on $TM$ can be pulled back to $M$. By taking into account that $\hat{g}_{IJ} \circ Y$ is a function defined on $U$, we can introduce a new metric, defined as
\begin{equation}\label{Y-riem}
 \hat{g}_{IJ}(x):=\hat{g}_{IJ}(x,y)|_{y=Y(x)},\quad x\in U.
\end{equation}

{\it The pair $(U,\hat{g}_{IJ})$} {\it correspond to a Riemannian manifold}, while  $\hat{g}_{IJ}(x)$ represents the $Y$-{\it osculating Riemannian metric} corresponding to $(M,F)$.

The Christoffel symbols of the first kind are defined for the osculating Riemannian metric \eqref{Y-riem} as,
\begin{eqnarray}
  \hat{\gamma}_{IJK}(x)&:=&\frac{1}{2}\left\{\frac{\partial}{\partial x^J}\left[\hat{g}_{IK}(x,Y(x))\right]
+\frac{\partial}{\partial x^K}\left[\hat{g}_{IJ}(x,Y(x))\right]\right.\nonumber\\
&&\left.-\frac{\partial}{\partial x^I}\left[\hat{g}_{JK}(x,Y(x))\right]\right\}.
\end{eqnarray}

Explicitly, after using the law of the derivative of the composed functions, we find
\bea\label{Christ for g_Y}
\hspace{-0.7cm}&&\hat{\gamma}_{IJK}(x)=\left.\hat{\gamma}_{IJK}(x,y)\right|_{y=Y(x)}\nonumber\\
\hspace{-0.7cm}&& +2 \left.\left(\hat{C}_{IJL}\frac{\partial Y^L}{\partial x^K}+\hat{C}_{IKL}\frac{\partial Y^L}{\partial x^J}-\hat{C}_{JKL}\frac{\partial Y^L}{\partial x^I} \right)\right|_{y=Y(x)},
\eea
where $\hat{C}_{IJL}$ is the Cartan tensor. Therefore, if a non-vanishing global section $Y$ of $TM$ does exist, with the property $Y(x)\neq 0$,  $\forall x\in M$, the osculating Riemannian manifold $(M,\hat{g}_{ij})$ can always be defined.

Let's now consider the case of an $(\alpha,\beta)$ metric. For this geometry we choose the vector field $Y=A$, with $A^{I}=g^{IJ}A_{J}$. By taking into account that the vector field $A$ is non-vanishing globally on $M$, it follows that $\beta $ has no zero points. Therefore, for $(\alpha, \beta)$ metrics we can define {\it the} $A$-{\it osculating Riemannian manifold} $(M,\hat{g}_{IJ})$, where the Riemann metric is given by $\hat{g}_{IJ}(x):=\hat{g}_{IJ}(x,A)$.
%, $A^{i}=a^{ij}A_{j}$.

For the length $\tilde{a}$ of $A$ with respect to $\alpha $ we immediately obtain  $\tilde{a}^{2}=A_{I}A^{I}=\alpha ^{2}\left( x,A\right)$ and  $Y_{I}\left( x,A\right)
=A_{I}$, respectively.

Explicitly, the $A$-osculating Riemannian metric can be written as
\begin{equation}
\begin{split}
   \hat{g}_{IJ}\left( x\right) & =\left.\frac{L_{\alpha }}{\tilde{a}}\right|_{y=A(x)}g_{IJ}\\
   & +\left.\left(
\frac{L_{\alpha \alpha }}{\tilde{a}^{2}}+2\frac{L_{\alpha \beta }}{\tilde{a}}%
+L_{\beta \beta }-\frac{L_{\alpha }}{\tilde{a}^{3}}\right)\right|_{y=A(x)} A_{I}A_{J}.
\end{split}
\end{equation}

Furthermore, we have the relations $\beta \left( x,A\right) =\tilde{a}^{2}$, and  $p_{I}\left(x,A\right) =0$, Therefore, from the definition of the Cartan tensor, and its expression for an $(\alpha,\beta)$ metric,  we obtain the basic result that $\hat{C}_{IJK}\left( x,A\right) =0$.
On the other hand, for  $Y=A$, we obtain
\be
\hat{\gamma}_{IJK}(x)=\left.\hat{\gamma}_{IJK}(x,y)\right|_{y=A(x)}.
\ee

Hence, we obtain the fundamental result that for {\it an $(\alpha ,\beta )$-metric, the Barthel connection - the linear $A$-connection with} $A^I=\left(g^{IJ}A_{J}\right)$, {\it is nothing but the Levi-Civita connection of the} $A${\it -Riemannian metric}. After evaluating the fundamental tensor $g_{ij}(x,y)$ of $(M,F)$ at $(x,Y(x))$, one obtains a Riemannian metric $g_Y$ on $M$, having its own Levi-Civita connection.

\subsection{The generalized curvature tensor}

As we have already seen, the Barthel connection, having local coefficients $\left(b_{BC}^A(x)\right)$, {\it is an affine connection}. The curvature tensor of an affine connection, with local coefficients $\left(\Gamma _{BC}^A(x)\right)$, is generally given by
\begin{equation}
R^A_{BCD}=\dfrac{\partial \Gamma^A_{BD}}{\partial x^C}-
    \dfrac{\partial \Gamma^A_{BC}}{\partial x^D}+\Gamma^E_{BD}\Gamma^A_{EC}
    -\Gamma^E_{BC}\Gamma^A_{ED}.
    \end{equation}

Hence, the curvature of the Barthel connection can be obtained from the above equation by taking $\left(\Gamma _{BC}^A(x)\right)=\left(b_{BC}^A(x)\right)$. In the case of the Kropina metric $F=\alpha^2/\beta$, on which we will focus in the following, the Barthel connection is identical with the Levi-Civita connection of the osculating metric $\hat{g}_{AB}(x)=g_{AB}\left(x,A(x)\right)$, where $A_I(x)$ are the components of the one-form $\beta$, and $g_{AB}$ is the fundamental tensor of $F$. Therefore, by taking into account that $b_{BC}^A=\hat{\gamma}_{BC}^A$, where $\hat{\gamma}_{BC}^A$ are the Levi-Civita connection coefficients, we obtain for the curvature tensors of the Kropina metric the expressions
\begin{equation}
\hat{R}^A_{BCD}=\dfrac{\partial \hat{\gamma}^A_{BD}}{\partial x^C}-
    \dfrac{\partial \hat{\gamma}^A_{BC}}{\partial x^D}+\hat{\gamma}^E_{BD}\hat{\gamma}^A_{EC}
    -\hat{\gamma}^E_{BC}\hat{\gamma}^A_{ED},
    \end{equation}
    and
    \begin{equation}
\hat{R}_{BD}=
    \displaystyle\sum_A\left[\dfrac{\partial \hat{\gamma}^A_{BD}}{\partial x^A}-\dfrac{\partial \hat{\gamma}^A_{BA}}{\partial x^D}
    +\sum _E\left(\hat{\gamma}^E_{BD}\hat{\gamma}^A_{EA}-\hat{\gamma}^E_{BA}\hat{\gamma}^A_{ED}\right)\right],
 \end{equation}
respectively, where the indices $A,B,C,D,E$ takes values in the set $\{0,1,2,3\}$, $\hat{R}_{BD}=\hat{R}_{BAD}^{A}$, and $\hat{R}_{D}^{B}=\hat{g}^{BC}\hat{R}_{CD}$, respectively \cite{Fc25}. Finally, the generalized Ricci scalar is defined as
$\hat{R}=\hat{R}_{B}^{B}$.

\section{Review of the Barthel-Kropina cosmological model}\label{sect2}

In the present Section we will review the basics and the background evolution of the Barthel-Kropina cosmological model, as introduced in \cite{Fc26}. The generalized Barthel-Kropina-Friedmann equations will represent the theoretical basis for the detailed comparison of this cosmological model with the observations. Moreover, they offer a deeper insight into the mathematical and physical structure of the model, and into the possible relevance of the Finsler geometric structure for the understanding and description of the cosmological dynamics.

\subsection{Metric and thermodynamic quantities}

The Barthel-Kropina cosmological model is based on a Finsler type $(\alpha, \beta)$ geometry, in which the fundamental metric function of the geometry is defined as $F=\alpha ^2/\beta$, with $\alpha$ a positive non-degenerate Riemann metric, and $\beta$ an one form. In the following we adopt a coordinate system with $\left(x^0=ct,x^1=x,x^2=y,x^3=z\right)$.

To build-up consistently a cosmological model based on the Kropina metric function, we need to supplement the Finsler framework with several mathematical and physical assumptions \cite{Fc26}, which we list below:

\paragraph {The Riemann metric $g_{IJ}(x)$ is Friedmann-Lemaitre-Robertson-Walker.} We postulate that the metric in $\alpha$
is given by the Friedmann-Lemaitre-Robertson-Walker (FLRW) metric,
\be
ds^2=\left(dx^0\right)^2-a^2\left(x^0\right)\left[\left(dx^1\right)^2+\left(dx^2\right)^2+\left(dx^3\right)^2\right],
\ee
where $a\left(x^0\right)$ is the scale factor. Hence, as assume that the Universe is homogeneous and isotropic, with a uniform flow of the cosmological time $t$.

\paragraph{Validity of the Cosmological Principle.} We assume the validity of the Cosmological Principle throughout the large scale Universe, by requiring that all geometrical and physical quantities depend on the cosmological time only. Thus, in the one-form $\beta$, the components of the vector $A$ depend on the cosmological time only, $A_I=A_I\left(x^0\right)$.

\paragraph{The vector $A$ has only one time-like independent component, $A_0\left(x^0\right)$.} We assume that the space-like components of $A$ vanish, so that $A_1=A_2=A_3=0$. The non-satisfaction of this condition leads to the violation of the Cosmological Principle, and to the existence of a preferred direction in the Universe.
But this would contradict the observationally well confirmed large scale spatial isotropy of the
Universe. Therefore, in the following we consider that the 1-form field $\beta $ has the form
\begin{align}\label{special A_i}
(A_{I})=(a\left(x^0\right)\eta\left(x^0\right),0,0,0)=(A^{I}),
\end{align}
where by $\eta \left(x^0\right)$ we have denoted an arbitrary function of the cosmological time.

\paragraph{Matter comoves with the cosmological expansion.} We assume that in the Barthel-Kropina geometry {\it a comoving frame}, in which all observers move together with the Hubble flow in the Riemannian geometry described by the metric $g_{IJ}(x)$.

\paragraph{Matter and thermodynamics.} We assume that the matter content of the Universe can be described as a perfect fluid, characterized by two thermodynamic quantities only, the energy density $\rho c^2$, and the thermodynamic pressure $p$, respectively. Since the matter is comoving with the cosmological expansion, and the Universe is homogeneous and isotropic, the only non-vanishing components of the matter energy-momentum tensor $\hat{T}_{AB}$ are
\be
 \hat{T}_0^0=\rho c^2, \hat{T}_{00}=\hat{g}_{00}\hat{T}^0_0, \hat{T}_k^k=-p, \hat{T}_{ii}=-\hat{g}_{ik}\hat{T}^k_i.
 \ee

\paragraph{Gravity from geometry.} We postulate that the Einstein gravitational field equations are given in the Barthel-Kropina geometry by
\begin{equation}\label{Eineq}
\hat{R}_{BD}-\frac{1}{2}\hat{g}_{BD}\hat{R}=\kappa ^2 \hat{T}_{BD},
\end{equation}%
where $G$ and $c$ are the Newtonian gravitational constant, and the speed of light, respectively, and $\kappa^2=8\pi G/c^4 $ is the gravitational coupling constant. $\hat{T}_{BD}$ denotes the matter energy-momentum tensor, obtained in the usual way from the standard thermodynamic quantities, with the help of the Finslerian metric tensor $\hat{g}_{BD}$.

Therefore, from the previous assumptions, we can summarize the mathematical structure  of the Barthel-Kropina-FLRW cosmological model as follows \cite{Fc26},
\begin{enumerate}[(i)]
%\item $\epsilon=1$;
\item $\left(A_I\right)=\left(a\left(x^0\right)\eta\left(x^0\right),0,0,0\right)=\left(A^I\right)$;
\item $\left(g_{IJ}\right)=\begin{pmatrix}
1 & 0 & 0 & 0 \\
0 & -a^2(x^0) & 0 & 0 \\
0 & 0 & -a^2(x^0) & 0\\
0 & 0 & 0 & -a^2(x^0)
\end{pmatrix};$
\item $\alpha|_{y=A(x)}=a(x^0)\eta(x^0)$;
\item $\beta|_{y=A(x)}=[a(x^0)\eta(x^0)]^2$.
\end{enumerate}

\subsection{The generalized Friedmann equations for the background cosmological evolution}

The Einstein gravitational field equations in the Barthel-Kropina-FLRW geometry,
\be
\hat{G}_{00}=\frac{8\pi G}{c^4}\hat{g}_{00}\rho,\hat{G}_{ii}=-\frac{8\pi G}{c^4}\hat{g}_{ii}p,
\ee
gives the generalized Friedmann equations for the background cosmological evolution, which take the form \cite{Fc26},
\be\label{Fr1}
\frac{3(\eta')^2}{\eta^2}=\frac{8 \pi G}{c^2}\frac{1}{a^2\eta ^2}\rho,
\ee
and
\be\label{Fr2}
2\frac{\eta''}{\eta}+2\cH\frac{\eta'}{\eta}-3\frac{(\eta')^2}{\eta ^2}=\frac{8\pi G}{c^4}\frac{p}{a^2\eta ^2},
\ee
respectively, where we have denoted $\cH=\left(1/a\left(x^0\right)\right)\left(da\left(x^0\right)/dx^0\right)$. By eliminating the term $-3\left(\eta '\right)^2$ by using Eq.~(\ref{Fr1}), Eq.~(\ref{Fr2}) takes the simple form
\be\label{Fr3}
a\eta \frac{d}{dx^0}\left(\eta 'a\right)=\frac{4\pi G}{c^4}\left(\rho c^2+p\right).
\ee

The set of the generalized Friedmann equations in the Barthel-Kropina geometry consists of two equations with four unknowns $\left(a,\eta, \rho, p\right)$. Even after imposing an equation of state for the ordinary matter, $p=p(\rho)$, the system of generalized Friedmann equations is still underdeteremined. Hence, to close it, we need to impose an independent relation between two of the model parameters.

\paragraph{Conservation of matter and energy.} In standard Friedmann cosmology the matter energy-density is conserved. This is not the case in the Barthel-Kropina-FLRW model. The matter energy-momentum balance equation can be obtained easily by multiplying Eq.~(\ref{Fr1}) with $a^3$, and taking the derivative of the result with respect to $x^0$. By combining this relation with the second generalized Friedmann equation, we find the energy balance equation in the Barthel-Kropina-FLRW cosmology as given by \cite{Fc26},
\bea\label{bal}
&&\frac{8 \pi G}{c^4}\left[\frac{d}{dx^0}\left(\rho c^2a^3\right)+p\frac{d}{dx^0}a^3\right]=6a^5\Big[\cH \left(\eta '\right)^2\nonumber\\
&&+\left(\eta '+\cH \eta\right)\eta ''+\cH ^2\eta \eta '\Big].
\eea

We can also rewrite the energy balance equation  in a form that is closer to the standard general relativistic result as
\bea
&&\frac{4 \pi G}{c^4}\left[\frac{d}{dx^0}\left(\rho c^2a^3\right)+p\frac{d}{dx^0}a^3\right]\nonumber\\
&&=3a^5\left[\frac{8\pi G}{2c^4}\left(\frac{5}{3}\rho c^2+p\right)\frac{\cH}{a^2}+\eta ' \eta ''\right].
\eea

\paragraph{Recovering standard General Relativistic cosmology.} An important characteristics of the cosmological equations of the Barthel-Kropina-FLRW model, Eqs.~(\ref{Fr1}) and (\ref{Fr2}), respectively,  is that they reduce to the standard Friedmann equations of general relativity {\it in the limit} $\eta \rightarrow \pm 1/a$, $\beta =(1,0,0,0)$,
\be
\frac{3(a')^2}{a^2}=\frac{8 \pi G}{c^2}\rho,\;\;
2\frac{a ''}{a}+\frac{(a')^2}{a^2}=-\frac{8\pi G}{c^4}p.
\ee

 The Friedmann equations of standard general relativistic cosmology lead to the conservation of the energy density $\rho$, which takes the form,
$\dot{\rho}+3H\left(\rho +p/c^2\right)=0$, where by a dot we have denoted the derivative with respect to the cosmological time $t$, and we have introduced the standard Hubble function defined as $H=c\cH$.

\section{The Barthel-Kropina dark energy model}\label{sect3}

In the present Section we will introduce the Barthel-Kropina dark energy model, which is based on an appropriate splitting of the generalized Friedmann equations (\ref{Fr1}) and (\ref{Fr2}).  More exactly, after introducing an appropriate representation for the coefficient $\eta$ of the one-form $\beta$, which brings the cosmological field equations to a form similar to the standard Friedmann equations, we will interpret the extra terms in these equations as describing a dynamical cosmological dark energy, whose effective energy density and pressure can be related by a linear equation of state.

\subsection{Alternative representation of the generalized Friedmann equation}

In order to bring the Barthel-Kropina-FLRW cosmological equations to a form as close as possible to the Friedmann equations of general relativity, we represent the function $\eta$ in a general form as
\be
\eta \left(x^0\right) =\frac{1}{a\left(x^0\right)}\left[1+\psi \left(x^0\right)\right],
\ee
where $\psi\left(x^0\right)$ is a time dependent arbitrary function, which must be determined from the field equations. For this form of $\eta$, the generalized Friedmann equations of the Barthel-Kropina-FLRW model take the form \cite{Fc26},
\bea\label{52}
3\frac{\left(a'\right)^2}{a^2}&=&\frac{8\pi G}{c^2}\rho+6(1+\psi) \psi ' \cH -3\left(\psi '\right)^2-3(2+\psi)\psi\cH ^2 \nonumber\\
&=&\frac{8\pi G}{c^2}\rho +\rho _{DE},
\eea
and
\bea\label{53}
\frac{2a''}{a}+\frac{\left(a'\right)^2}{a^2}&=&-\frac{8\pi G}{c^4}\frac{p}{(1+\psi)^2}+4 \frac{\psi '}{1+\psi}\cH-3\frac{\left(\psi '\right)^2}{(1+\psi)^2}\nonumber\\
&&+2\frac{\psi ''}{1+\psi}=-\frac{8\pi G}{c^4}\frac{p}{(1+\psi )^2}+p_{DE},
\eea
respectively, where we have denoted
\be
\rho_{DE}=6(1+\psi) \psi '\cH -3\left(\psi '\right)^2-3(2+\psi )\psi \cH ^2,
\ee
and
\bea
p_{DE}=4 \frac{\psi '}{1+\psi}\cH-3\frac{\left(\psi '\right)^2}{(1+\psi)^2}+2\frac{\psi ''}{1+\psi},
\eea
respectively. For $\psi \rightarrow 0$, and $\eta \rightarrow 1/a$, we reobtain the Friedmann equations of standard cosmology. The generalized Friedmann equations describe the evolution of the Universe with the standard general relativistic evolution modified by the presence of extra terms generated by the Finsler geometric effects in the cosmological space-time.

\subsection{Dark energy, and its equation of state}

 We interpret the extra terms in Eqs.~(\ref{52}) and (\ref{53}) as describing {\it an effective, dynamical geometrical fluid type dark energy}, with energy density $\rho_{DE}$, and pressure $p_{DE}$, respectively. To obtain an effective dark energy term,  we impose on the geometric fluid the equation of state
 \be
 p_{DE}=\omega \rho_{DE}, \omega ={\rm constant},
 \ee
 which gives for the  function $\psi $ the following differential equation,
\bea\label{47}
&&\frac{2 \psi ''}{1+\psi}+2 \cH \left[-3 \omega (1+\psi)
   +\frac{2}{1+\psi}\right] \psi ' \nonumber\\
  && +3\left[
   \omega-\frac{1}{(1+\psi)^2}\right] \left(\psi '\right)^2+3 \omega \psi (2+\psi)
   \cH ^2=0.
\eea

The energy density and the pressure of the dynamical cosmological dark energy are dependent on the scale factor, and on the properties of coefficient $\eta$ of the one form $\beta$.

In the limit of small values of $\psi\left(x^0\right)$, $\psi \left(x^0\right)<<1$, corresponding to small deviations from standard general relativity, and in the de Sitter limit, with $\cH=\cH_0={\rm constant}$,  Eq.~(\ref{47}) can be written as
\be\label{48}
\psi ''+\cH_0(2-3\omega)\psi '+3\cH_0\omega \psi=0,
\ee
where we have also neglected the small term containing $\left(psi '\right)^2$. Eq.~(\ref{48}) has the general solution
\be
\psi \left(x^0\right)=e^{(3\omega-2)\cH_0x^0}\left(C_1e^{\sqrt{\delta}\cH_0x^0}+C_2e^{-\sqrt{\delta}\cH_0x^0}\right),
\ee
where $C_1$ and $C_2$ are arbitrary constants of integration, and $\delta =9\omega ^2-24\omega +4$. $delta$ is negative for $\omega $ in the range $\omega \in (0.1,2.5)$, and hence for these values of the equation of state parameter, and in the de Sitter regime, $psi$ has an oscillatory behavior. For small values of $\omega$, so that $\omega <<2/3$, $\psi$ becomes a constant, $\psi =C_1$.

We will consider in the following only the late time cosmological evolution of the Universe, and hence we will assume that the ordinary matter pressure vanishes, $p=0$.

\subsection{Cosmological evolution in the redshift space}

In order to simplify the mathematical formalism, and to bring it closer to the observational data,  we introduce the dimensionless time parameter $\tau$, defined as $\tau=\cH _0x^0$, as well as the normalized Hubble function $h$, defined as $\cH =\cH _0h$, with $\cH_0=H_0/c$, and $H_0$ denoting the present day value of the Hubble function. Furthermore, we denote $\sigma (\tau)=d\psi (\tau)/d\tau$.

Hence, the basic equations describing the time dynamics of the generalized Friedmann equations of the Barthel-Kropina-FLRW cosmological model take the form
\be\label{de1}
\frac{d\psi }{d\tau}=\sigma,
\ee
\be\label{de2}
2\frac{dh}{d\tau}+3h^2=4h\frac{\sigma}{1+\psi}-3\frac{\sigma ^2}{(1+\psi )^2}+\frac{2}{1+\psi}\frac{d\sigma}{d\tau},
\ee
\bea\label{de3}
&&\frac{2}{1+\psi}\frac{d\sigma}{d\tau}+2h\left[\frac{2}{1+\psi}-3\omega (1+\psi)\right]\sigma\nonumber\\
&&+3\left[\omega -\frac{1}{(1+\psi)^2}\right]\sigma^2+3\omega \psi(2+\psi)h^2=0.
\eea

We also introduce the critical matter density $\rho _c=3H_0^2/8\pi G$, as well as the matter density parameter defined according to $\Omega _m=\rho /\rho _c$. Then,  from the first generalized Friedmann equation we obtain the matter density parameter in the form
\be
\Omega _m=h^2+\sigma ^2+(2+\psi)\psi h^2-2(1+\psi)h\sigma.
\ee

To allow a direct confrontation of the theoretical model with the observations, we will reformulate the cosmological Barthel-Kropina-FLRW evolution equations in terms of the redshift variable $z$, defined as $1+z=1/a$. Hence, in the redshift space the system of equations (\ref{de1})-(\ref{de3}) take the form
\be\label{deq1}
-(1+z)h\frac{d\psi}{dz}=\sigma,
\ee
\bea\label{deq2}
-2(1+z)h\frac{dh}{dz}+3h^2&=&4h\frac{\sigma}{1+\psi}-3\frac{\sigma ^2}{(1+\psi)^2}\nonumber\\
&&-2(1+z)\frac{h}{1+\psi}\frac{d\sigma}{dz},
\eea
\bea\label{deq3}
&&-2(1+z)\frac{h}{1+\psi}\frac{d\sigma}{dz}+2h\left[\frac{2}{1+\psi}-3\omega (1+\psi)\right]\sigma \nonumber\\
&&+3\left[\omega -\frac{1}{(1+\psi)^2}\right]\sigma ^2+3\omega \psi(2+\psi)h^2=0.
\eea

 The matter density parameter is obtained in the redshift space as
 \be
 \Omega _m=h^2\Bigg[1+(1+z)^2\left(\frac{d\psi}{dz}\right)^2+(2+\psi)\psi+2(1+z)(1+\psi)\frac{d\psi}{dz}\Bigg].
 \ee

The general properties of a cosmological model can be extracted from the study of a number of specific observational quantities. The first such quantity we will consider is the deceleration parameter, defined as
\be
q=\frac{d}{d\tau}\frac{1}{H}-1=-\frac{1}{a}\frac{d^2a}{d\tau ^2}\left[\frac{1}{a}\frac{da}{d\tau}\right]^{-2}.
\ee

 The deceleration parameter of the Barthel-Kropina-FLRW cosmological model can be obtained explicitly in the form
 \be
 q=\frac{1}{2}+\frac{3}{2}\frac{\frac{8\pi G}{c^4}\frac{p}{(1+\psi)^2}-p_{DE}}{\frac{8\pi G}{c^2}\rho +\rho_{DE}}.
 \ee

 We will also investigate the behavior of the jerk and snap parameters $j$ and $s$, defined according to
 \be
 j=\frac{1}{a}\frac{d^3a}{d\tau ^3}\left[\frac{1}{a}\frac{da}{d\tau}\right]^{-3}=q(2q+1)+(1+z)\frac{dq}{dz},
 \ee
 and
 \be
 s=\frac{1}{a}\frac{d^4a}{d\tau ^4}\left[\frac{1}{a}\frac{da}{d\tau}\right]^{-4}=\frac{j-1}{3\left(q-\frac{1}{2}\right)},
 \ee
 respectively.

\section{Cosmological tests of the Barthel-Kropina dark energy model}\label{sect4}

In the present Section we will perform a detailed comparison between the predictions of the Barthel-Kropina-FLRW dark energy model, and we will also investigate the observational constraints on the model. In our analysis we are going to use two observational datasets, the $H(z)$ data,  and the Pantheon data set, which contain 57, and 1048 points, respectively. From the comparison with the cosmological data we find the best fit ranges of the four free parameters of the Barthel-Kropina dark energy model, $\left(\omega, \sigma_{0}, \psi_{0}\right)$, which must be considered together with the present day value of the Hubble function $h$. To constrain the cosmological model parameters,  we use the standard Bayesian technique, and likelihood function approach, along with the Markov Chain Monte Carlo (MCMC) method. Once the best fit values of the model parameters are known, we will investigate the model from a cosmographic point of view, by analysing the evolution of the deceleration, jerk, and snap parameters. In all cases we compare the model predictions with the standard $\Lambda$CDM cosmological model.

\subsection{Methodology}

Presently, a large amount of observational data, obtained from different cosmological observations,  such as Supernovae type Ia, Baryon Acoustic oscillations (BAO), Cosmic Microwave Background (CMB), and Hubble rate measurements, respectively, can be used to test cosmological theories. This comparison between observation and theory allows to clearly discriminate between observationally supported models, and the unsupported ones. Due to the presence of a large amount of data, statistical analysis and methods  are necessary to confront theoretical models and observational data. The estimation of the cosmological parameters is performed in the framework of the Bayesian inference \cite{Trotta:2008qt,Liddle:2009xe}. In this approach the model parameters are considered as random variables, described by probability distributions. In the Bayesians interpretation probabilities are considered as a degree of belief in a hypothesis. Mathematically,  this is expressed by Bayes' theorem,
\begin{equation}
\mathcal{P}(H|D)=\frac{\mathcal{L}(D|H)\mathcal{P}(H)}{\mathcal{P}(D)},
\end{equation}
where $D$ denotes the observational/experimental data, while $H$ denotes the hypothesis to be tested, which usually can be considered as the parameters of the model, or  the model itself. Moreover, $\mathcal{P}(H|D)$, $\mathcal{P}(H)$, and $\mathcal{P}(D)$, respectively,  denote the posterior distribution, the prior distribution, and the evidence.  The later is a normalization factor, often omitted, since it is model independent. On the other hand, $\mathcal{P}(H)$ is taken mainly from previous experiments, and acquired information about the parameters. For a flat prior, i.e., for $\mathcal{P}(H)=1$,  Bayes' theorem takes the form,
\begin{equation}
\mathcal{P}(H|D)\propto \mathcal{L}(D|H).
\end{equation}

Although Bayes' theorem seems very simple, it relates  the posterior distribution  $\mathcal{P}(H|D)$ to the easily computable quantity, $\mathcal{L}(D|H)$, which expresses the probability of the hypothesis being true. In the case where the errors follow a  Gaussian distribution, the likelihood function takes the form,
\begin{equation}\label{chi2_L}
\chi^2 (\theta)=-2\ln{\mathcal{L}}(\theta).
\end{equation}

To estimate the model parameters, we look for the parameters vector  maximizing  the posterior distribution, i.e., $\hat{\theta}_{\textrm{b.f}}$ \footnote{We denote the best fit parameters, or the best fit vector, by $b.f$}.  From Eq.~(\ref{chi2_L}), one can see that maximizing the posterior distribution  is practically the same as minimizing the chi-squared function. The above results indicate the theoretical framework of the parameters inference.

In order to infer the best fit values of the cosmological parameters we use the Markov Chains Monte Carlo (MCMC) method \cite{Nesseris:2017vor}. In what follows, we  give a short description to the basic concepts of the MCMC approach, taking as an example the Metropolis-Hasting algorithm \cite{Metropolis-Hasting}. We summarize the MCMC algorithm used in the present study in five steps:\\

a) Choose a  starting point, which could be just a guess, or it could be  based on a previous estimates, or theoretical inferences. \\

b) Generate a new point from a proposed initial distribution, which is in general a multivariate normal distribution.\\

c) Compare the posterior height  of the new proposal to the posterior height of the current points. If the new proposal has a higher posterior value than the most current state, then accept the new proposal. If not,  choose to accept, or reject randomly, the new proposal. \\

d) Attribute the new proposal  to the  next sample, if it is accepted, otherwise  copy the most recent sample to the next sample.\\

e) Repeat the above steps, except initialization, until running all the iterations.   \\

%%%%%%%%%%%
To graphically present the errors in two dimensions, as well as the correlations between the estimated model parameters, it is convenient to plot the iso-probability contours,
\begin{equation}
\chi^2=\chi^2_{min}+\Delta\chi^2(\nu,n\sigma),
\end{equation}
where  $\chi^2_{min}=\chi^2(\hat{\theta}_{\textrm{b.f}})$ is the minimal chi-square value, and $\Delta\chi^2$ is the desired confidence level, calculated as follows,
\begin{equation}
\Delta\chi^2(\nu,n\sigma)=2\mathcal{G}\left(\frac{\nu}{2},1-{\rm erf}\left(\frac{n\sigma}{\sqrt{2}}\right)\right) ,
\end{equation}
where $\mu$ and $n\sigma$ denote the number of free parameters, and the confidence interval, respectively, while  $\mathcal{G}$ is the inverse of the regularized $\Gamma (x)$ function. Moreover, ${\rm erf}(x)$ is the error function,  given by,
\begin{equation}
{\rm erf}(x)=\frac{2}{\sqrt{\pi}}\int_{0}^{x}e^{-t^2}dt.
\end{equation}

To figure out which model gives the best description of the observational data we use the corrected Akaike's Information Criterion ($AIC_c$), given by \cite{AIC},
\begin{equation}
AIC_{c} = \chi^2_{min}+2\mathcal{N_{\textrm{f}}}+\frac{2\mathcal{N_{\textrm{f}}}(\mathcal{N_{\textrm{f}}}+1)}{\mathcal{N_{\textrm{Tot}}}-\mathcal{N_{\textrm{f}}}-1},
\end{equation}
where $\mathcal{N_{\textrm{f}}}$ and $\mathcal{N_{\textrm{Tot}}}$ denote the number of free parameters, and  the total data points used in the observational confrontation, respectively.

The $AIC_c$ is a very powerful tool when it comes to classify  a set of models according to their statistical significance. By definition, the  model with  the minimal $AIC_c$  is the most supported, and taken to be a reference. In practice, we are interested in the quantity
\be
\Delta AIC_c =  AIC_{c,model}- AIC_{c,reference},
\ee
which reveals how much each model is statistically close to the  reference model. If $0 < \Delta \textrm{AIC}_c< 2$, the model is substantially supported, and if  $4 < \Delta \textrm{AIC}_c< 7$, the model has less observational support. Models with $\Delta \textrm{AIC}_c> 10$  are not supported with respect to the reference model.

\subsection{Data description}

In the present Section we briefly present the observational datasets used for the tests of the Barthel-Kropina cosmological model.

\subsubsection{$H(z)$ Dataset}

In order to obtain strong constraints on the cosmological model parameters  one must confront it with several observational datasets. In our analysis of the Barthel-Kropina dark energy model, to constrain the model parameters, we make use of the  $H(z)$ measurements, together with the Pantheon sample. In general, Hubble data can be derived by measuring the BAO in the radial direction of galaxy clustering \cite{Gaztanaga:2008xz}, or by the differential age approach, in which the redshift dependence of the Hubble function is given by,
$$
H(z)=-\frac{1}{1+z} \frac{d z}{d t},
$$
where ${d z}$/ ${d t}$ is inferred from  two passively evolving galaxies. In our analysis, 57 data points for $H(z)$, distributed in the redshift range $0.07 \leqslant z \leqslant 2.42$.\\

To estimate the model parameters $\omega, \sigma_{0}, \psi_{0}$ and $h$, we compare the theoretical expectations of the Barthel-Kropina  model with  the uncorrelated Hubble measurements, using the chi-square function, defined as
\begin{equation}
\chi_{H}^{2}\left(\omega, \sigma_{0}, \psi_{0}, h\right)\\=\sum_{i=1}^{57} \frac{\left[H_{t h}\left(z_{i}, \omega, \sigma_{0}, \psi_{0}, h\right)-H_{o b s}\left(z_{i}\right)\right]^{2}}{\sigma_{H\left(z_{i}\right)}^{2}},
\end{equation}
where $H_{\text {th }}$, $H_{o b s}$ and $\sigma_{H\left(z_{i}\right)}$  represent the model prediction, the observed value of Hubble rate, and  the standard error at the redshift $z_{i}$, respectively. The numerical values of the Hubble function at the corresponding redshifts are presented in Table~\ref{T1}.

\begin{table}
\begin{center}
\begin{tabular}{|c|c|c|c|c|c|c|c|}
%\toprule\toprule
\multicolumn{8}{c}{\bf Observational Hubble  parameter data } \\
%\toprule
\hline
       $z$ & $H(z)$ & $\sigma_H$  & Ref.                    & $z$     & $H(z)$   & $\sigma_H$ & Ref. \\
\hline $0.070$  & $69$ & $19.6$ & [146]       & $0.51$ & $90.4$    & $1.9$    & [152]\\[0.1cm]
\hline $0.120$  & $68.6$ & $26.2$ & [146]    & $0.52$ & $94.35$  & $2.64$    & [149]\\[0.1cm]
\hline $0.170$  & $83$ & $8$ & [147]         & $0.56$  & $93.34$  & $2.3$    & [149]\\[0.1cm]
\hline $0.1791$  & $75$ & $5$ & [150]         & $0.57$  & $87.6$   & $7.8$    & [155]\\[0.1cm]
\hline $0.1993$  & $75$ & $5$ & [150]          & $0.57$  & $96.8$  & $3.4$    & [156]\\[0.1cm]
\hline $0.200$  & $72.9$ & $29.6$ & [154]         & $0.59$  & $98.48$   & $3.18$    & [149]\\[0.1cm]
\hline $0.240$  & $79.69$ & $2.99$ & [139]         & $0.593$  & $104$  & $13$    & [150]\\[0.1cm]
\hline $0.270$  & $77$ & $14$ & [147]        & $0.60$  & $87.9$  & $6.1$    & [153]\\[0.1cm]
\hline $0.280$   & $88.8$ & $26.6$ & [154]          & $0.61$  & $97.3$  & $2.1$    & [152]\\[0.1cm]
\hline $0.300$  & $81.7$ & $6.22$ & [148]        & $0.64$  & $98.82$  & $2.98$    & [149]\\[0.1cm]
\hline $0.310$  & $78.18$ & $4.74$ & [149]        & $0.6797$  & $92$  & $8$    & [150]\\[0.1cm]
\hline $0.340$  & $83.8$ & $3.66$ & [139]        & $0.73$  & $97.3$  & $7$    & [153]\\[0.1cm]
\hline $0.350$  & $82.7$ & $9.1$ & [151]       & $0.7812$  & $105$  & $12$     & [150]\\[0.1cm]
\hline $0.3519$ & $83$ & $4$ & [150]       & $0.8754$  & $125$  & $17$    & [150]\\[0.1cm]
\hline $0.360$ & $79.94$ & $3.38$ & [149]       & $0.880$  & $90$  & $40$ & [146]\\[0.1cm]
\hline $0.380$ & $81.5$ & $1.9$ & [152]      & $0.900$  & $69$   & $12$     & [147]\\[0.1cm]
\hline $0.3802$  & $83$ & $13.5$ & [157]      & $0.900$  & $117$   & $23$     & [147]\\[0.1cm]
\hline $0.400$ & $95$ & $17$ & [147]       & $1.037$  & $154$   & $20$     & [150]\\[0.1cm]
\hline $0.400$ & $82.04$ & $2.03$ & [149]     & $1.300$  & $168$   & $17$     & [147]\\[0.1cm]
\hline $0.4004$ & $77$ & $10.2$ & [157]     & $1.363$  & $160$   & $33.6$     & [156]\\[0.1cm]
\hline $0.4247$  & $87.1$ & $11$ & [157]      & $1.430$  & $177$   & $18$     & [147]\\[0.1cm]
\hline $0.43$  & $86.45$ & $3.97$ & [139]     & $1.530$  & $140$   & $14$     & [147]\\[0.1cm]
\hline $0.44$  & $82.6$ & $7.8$ & [153]     & $1.750$  & $202$   & $40$     & [147]\\[0.1cm]
\hline $0.44$ & $84.81$ & $1.83$ & [149]   & $1.965$  & $186.5$   & $50.4$     & [156]\\[0.1cm]
\hline $0.4497$ & $92.8$ & $12.9$ & [157]   & $2.30$  & $224$    & $8.6$     & [159]\\[0.1cm]
\hline $0.470$ & $89$ & $34$ & [162] & $2.33$  & $224$    & $8$       & [160]\\[0.1cm]
\hline $0.4783$ & $80$ & $99$ & [157]   & $2.34$  & $222$    & $8.5$   & [159]\\[0.1cm]
\hline $0.480$ & $87.79$ & $2.30$ & [149]       & $2.36$  & $226$    & $9.3$   & [161]\\[0.1cm]
\hline $0.480$  & $97$ & $62$ & [146]   & & & & \\
%\toprule\toprule
\hline
\end{tabular}
\caption{The 57 data points of the Hubble  parameter considered in the present paper.}\label{T1}
\end{center}
\end{table}

\subsubsection{Pantheon Dataset}

The observation of type Ia supernova (SN Ia) had played a key role in the discovery of the comic accelerating expansion. Thus far, SN Ia are one of the most effective tools for investigating  the nature of the dark energy. During the recent  years, various supernova data compilations  have been released. We cite for instance Union \cite{SupernovaCosmologyProject:2008ojh} , Union2 \cite{Amanullah:2010vv} and  Union2.1 \cite{SupernovaCosmologyProject:2011ycw} from  the Supernova Cosmology Project, Joint Light-cure Analysis (JLA) \cite{SDSS:2014iwm} and the Pantheon data \footnote{An update of the Pantheon sample was  recently released \cite{Scolnic:2021amr}. } \cite{Pan-STARRS1:2017jku}. The later, contains 1048 spectroscopically confirmed  SN Ia covering the  redshift range $0<z< 2.3$. Besides, SN Ia are astrophysical objects,  considered as standard candles, which  measure relative distances. Therefore, SN Ia sample are used via the distance modulus $\mu=m-M$, where $m$ denotes the apparent magnitude of a given SN Ia.

The chi-square of the SN Ia dataset is given by
\begin{equation}
 {\chi}_{SN}^2={\Delta \mu}^{T}\hspace{0.1cm}.\hspace{0.1cm}{\bf C}_{SN}^{-1}\hspace{0.1cm}.\hspace{0.1cm}{\Delta \mu} .
\end{equation}
Here ${\bf C}_{SN}$ is the covariance matrix, and  ${\Delta \mu}=\mu_{obs}-\mu_{th}$, where $\mu_{obs}$ denotes the observed distance modulus of a given SN Ia, while   $\mu_{th}$ is the theoretical distance modulus, given by
\begin{equation}
\mu_{th}(z)=5\log_{10} \frac{D_L(z)}{(H_0/c) Mpc} +25,
\end{equation}
where $H_0$ is the Hubble rate at the present time, and  $c$ is the speed of light. For the flat, homogeneous and isotropic  Friedmann-Lema\^itre-Robertson-Walker (FLRW) Universe, the luminosity distance, $D_L$, is expressed as follows,
\begin{equation}
D_L(z)=(1+z)H_0\int_{0}^{z}\frac{dz^{\prime}}{H\left(z^{\prime}\right)}.
\end{equation}

Since we also constrain the free parameters of the model, i.e., ${\omega, \sigma_0, \psi_0, h}$ by using the Pantheon sample, and therefore,
\begin{eqnarray}
 {\chi}_{SN}^2(\omega, \sigma_0, \psi_0, h)&=&{\Delta \mu}^{T}(\omega, \sigma_0, \psi_0, h)\times {\bf C}_{Pantheon}^{-1}\nonumber\\
&& \times {\Delta \mu}(\omega, \sigma_0, \psi_0, h) .
\end{eqnarray}

The total chi-square of the Barthel-Kropina dark energy model is therefore given by,
\begin{equation}
\chi^{2}_{tot}(\omega, \sigma_0, \psi_0, h)=\chi_{H}^{2}+{\chi}_{SN}^2.
\end{equation}

\subsection{Results}

To figure out if the  Barthel-Kropina dark energy model is observationally supported, and to determine the level of this support, we confront the model with the cosmological observations by minimizing the $\chi^2$ function. Since from the model we do not have an explicit form of the Hubble parameter, $H(z)$, the system of differential equations defining the Barthel-Kropina dark energy model, Eqs.~ ((\ref{de1})-(\ref{de3})), is solved numerically. In practice, to solve a given   differential equation one must define a set of initial conditions. For the Barthel-Kropina dark energy model, the initial conditions themselves are free parameters, to be obtained from observations.

Therefore, the system of differential equations must be  solved for each MCMC iteration, thus obtaining a numerical solution for $H(z)$ for each MCMC state. Once we have the numerical solution, we  compute the total chi-square  $\chi^{2}_{tot}(\omega, \sigma_0, \psi_0, h)$.

\subsubsection{Constraining the Barthel-Kropina dark energy model parameters}

In Table~\ref{tab_MCMC}, we present the best fit, and the mean values of the Barthel-Kropina dark energy model parameters,  with their corresponding errors. The  prior used in this analysis is also shown.

\begin{widetext}
\begin{table*}[htbp]
%\resizebox{\columnwidth}{!}
{\begin{tabular}{|c|c|c|c|c|}
\hline
%\toprule\toprule
{\bf Model} &{\bf Parameter} &{\bf Prior} &{\bf Best fit} &{\bf Mean} \\
\hline
%\toprule\toprule
$  \Lambda$CDM & $\Omega_{\mathrm{m}}$ &$[0.001,1]$ &$0.27859_{-0.0139588}^{+0.0139588}$   &$0.279249_{-0.0139493}^{+0.0139493}$ \\[0.1cm]
 &$h$ &$[0.4,1]$ &$0.691892_{-0.0088881}^{+0.008881}$       &$0.691857_{-0.00888348}^{+0.0088838}$ \\[0.1cm]
\hline
\multirow{3}{*}{ Barthel-Kropina} &$\omega$     &$[0,6]$ &$2.02382_{-0.819655}^{+0.819655}$ &$2.53314_{-0.817497}^{+0.817497}$ \\[0.1cm]
 &$\sigma_0$ &$[-3,3]$   &$0.619373_{-0.0939505}^{+0.0939505}$   &$0.652573_{-0.0936745}^{+0.0936745}$ \\[0.1cm]
 &$\psi_0$ &$[-3,3]$   &$0.40882_{-0.141627}^{+0.141627}$      &$0.458615_{-0.141176}^{+0.141176}$   \\[0.1cm]
 &$h$   &$[0.4,1]$  &$0.684579_{-0.0108078}^{+0.0108078}$   &$0.683052_{-0.0108017}^{+0.0108017}$ \\[0.1cm]
%\toprule
\hline
\end{tabular}}
\caption{ Summary of the best fit and of the mean values of the free cosmological parameters of the Barthel-Kropina dark energy model.}\label{tab_MCMC}
\end{table*}
%\end{center}
\end{widetext}

\subsubsection{Comparison with the $\Lambda$CDM model}

To test statistically the Barthel-Kropina dark energy model, we compare it with the standard $\Lambda$CDM cosmological model. For that purpose, we use the $AIC_c$ criterion, which takes into account the number of free parameters,  and the total number of data points.  To obtain a reliable result, we  constrain the $\Lambda$CDM and the Barthel-Kropina dark energy model with the same data type, i.e., $H(z)+{\rm Pantheon}$. The analysis shows that even that the Barthel-Kropina dark energy model is not the most favoured  ($A I C_c^{\Lambda \textrm{CDM}} < A I C_c^{Barthel-Kropina}$), it is still very well supported by observations ($\Delta A I C_c=0.48$) with respect to the most accepted $\Lambda$CDM model, making  the Barthel-Kropina dark energy model extremely competitive to the standard model of cosmology. The statistical details of the comparison of the $\Lambda$CDM model and of the Barthel-Kropina dark energy model are presented in Table~\ref{tab_AIC}.

%\begin{widetext}
\begin{table*}[htbp]
\begin{center}
%\resizebox{\columnwidth}{!}
{\begin{tabular}{|c|c|c|c|c|}
%\toprule\toprule
\hline
{\bf Model} & ${\chi_{\text{tot}}^2}^{min} $ & $\chi_{\text {red }}^2$ & $A I C_c$ & $\Delta A I C_c$ \\[0.1cm]
\hline
$\Lambda$CDM  &$1081.5479$ &$0.978776$ &$1085.56$ &0 \\[0.1cm]
\hline
 Barthel-Kropina    &$1078.0028$   &$0.975568$   &$1086.04$ &$0.48$ \\[0.1cm]
%\toprule
\hline
\end{tabular}}
\caption{Summary of the ${\chi_{\text{tot}}^2}^{min} $, $\chi_{\text {red }}^2$, $A I C_c$ and $\Delta A I C_c$.}\label{tab_AIC}
\end{center}
\end{table*}
%\end{widetext}

\begin{figure*}[htbp]
\includegraphics[scale=0.7]{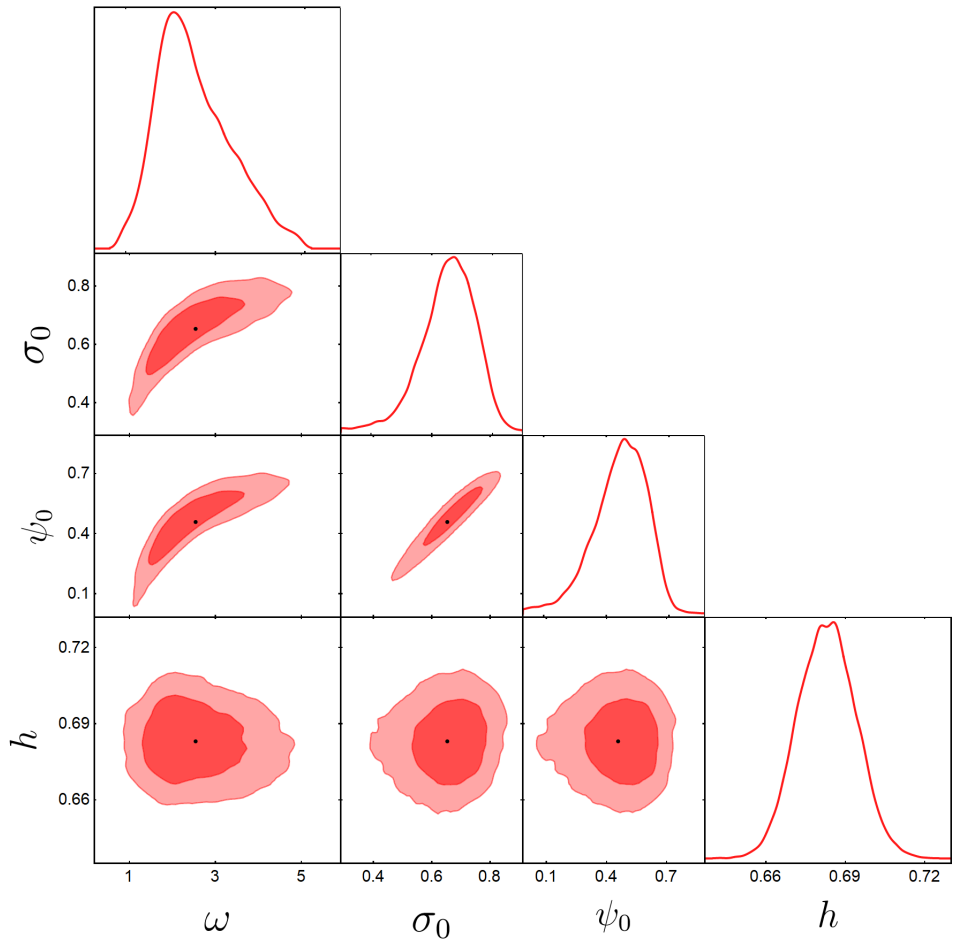}
\caption{MCMC confidence contours  at 1$\sigma$ and 2$\sigma$,  obtained after constraining the Barthel-Kropina dark energy model with SNIa+$\textrm{H}(z)$ data.}\label{fig_MCMCa}
\end{figure*}

\paragraph{Confidence levels.} In Fig.~\ref{fig_MCMCa},  the 1D and 2D posterior distributions at  $68.3\%$ (1$\sigma$) and $95.4\%$ (2$\sigma$) confidence levels are shown, obtained after constraining the Barthel-Kropina dark energy model with the Pantheon+$H(z)$ observational data. Besides, the MCMC contour plots show a  positive correlation between $(\omega, \sigma_0)$, $(\omega, \psi_0)$ and $(\sigma_0,\psi_0)$, while they shows a negative correlation between $(h,\omega)$ and $(h,\psi_0)$, respectively.
%\newpage
%\newpage

\subsubsection{Observational, and theoretical comparisons of the Hubble functions}

Now, once the free parameters of the Barthel-Kropina dark energy model are obtained, we can proceed to compare the model predictions with the observational data, and with the $\Lambda$CDM model, respectively.

\begin{figure}[htbp]
\includegraphics[scale=0.38]{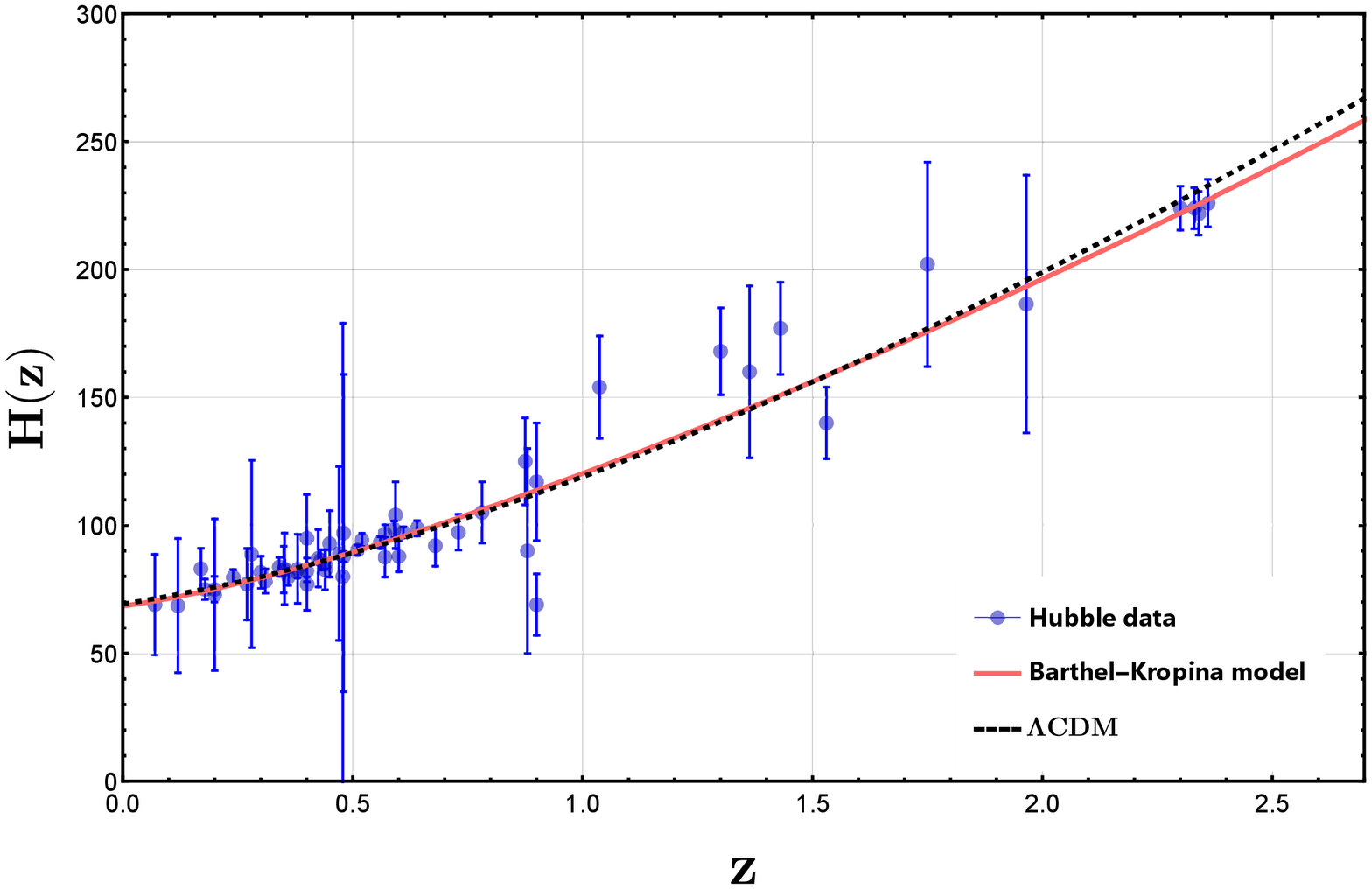}
\caption{The  evolution of the  Hubble parameter $\mathrm{H}(\mathrm{z})$ of the Barthel-Kropina and $\Lambda$CDM models as a function of the redshift $z$ against the Hubble measurements. }\label{HZ}
\end{figure}

\paragraph{Comparison with the Hubble data points.} First, we consider the comparison of the Barthel-Kropina dark energy model with the Hubble 57 data points, and with the $\Lambda$CDM model. The results of the comparison are presented in Fig.~\ref{HZ}. We observe from the Figure that the Barthel-Kropina dark energy model describes very well the Hubble data.

\begin{figure}[htbp]
\includegraphics[scale=0.43]{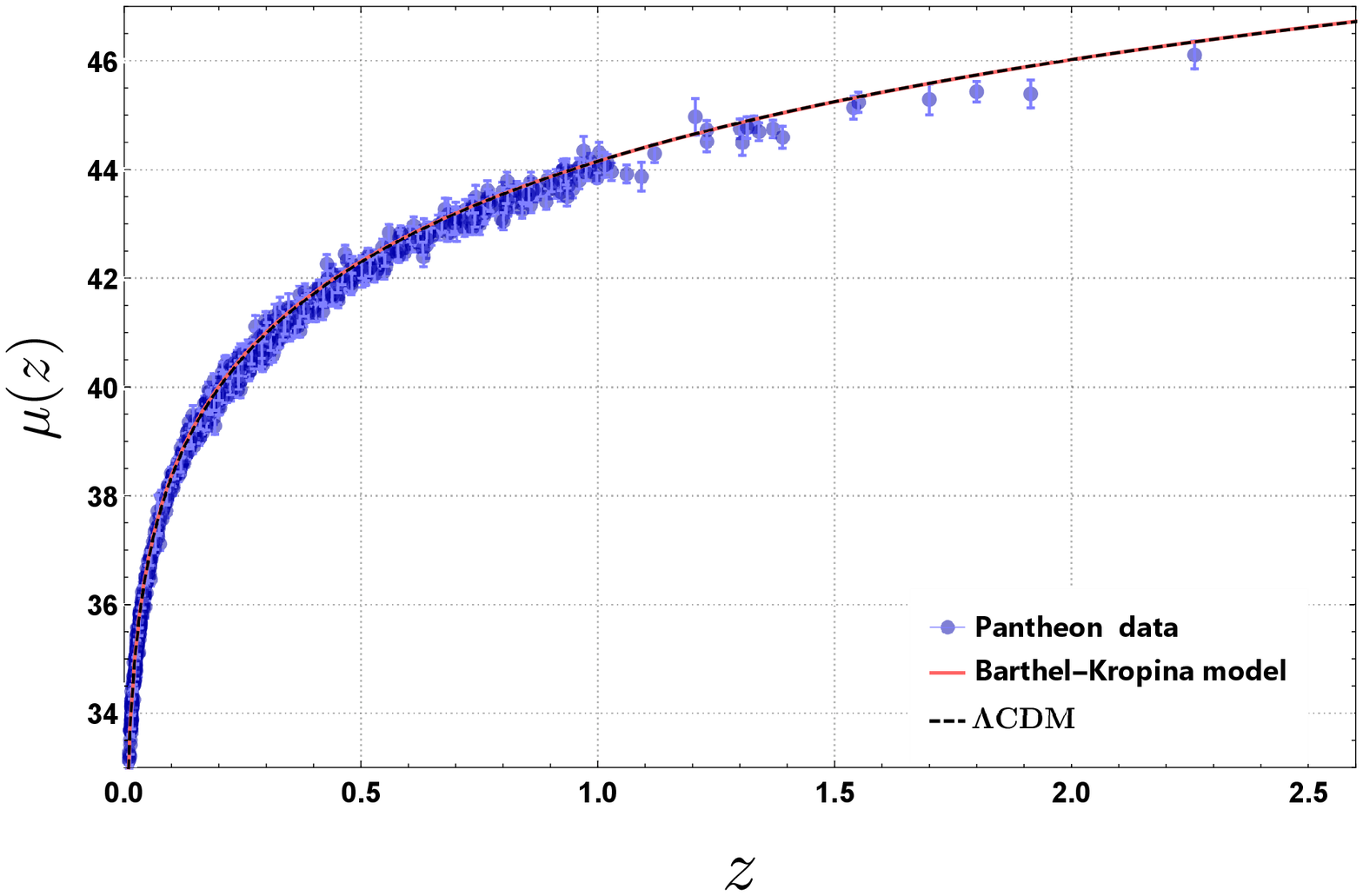}
\caption{The evolution of the distance modulus $\mu({z})$  of the Barthel-Kropina dark energy model, and of the $\Lambda$CDM model in terms of the redshift $z$ against the Pantheon data.}\label{mun}
\end{figure}

\paragraph{Comparison with the Pantheon data.} We proceed now to compare the  $\mu({z})$ distance modulus function  of the Barthel-Kropina dark energy model with the Pantheon data. From Fig.~\ref{mun}, one can see that  Barthel-Kropina model fits the Pantheon distance modulus, based on 1048 observation points,  very well.

\begin{figure}[htbp]
\includegraphics[scale=0.42]{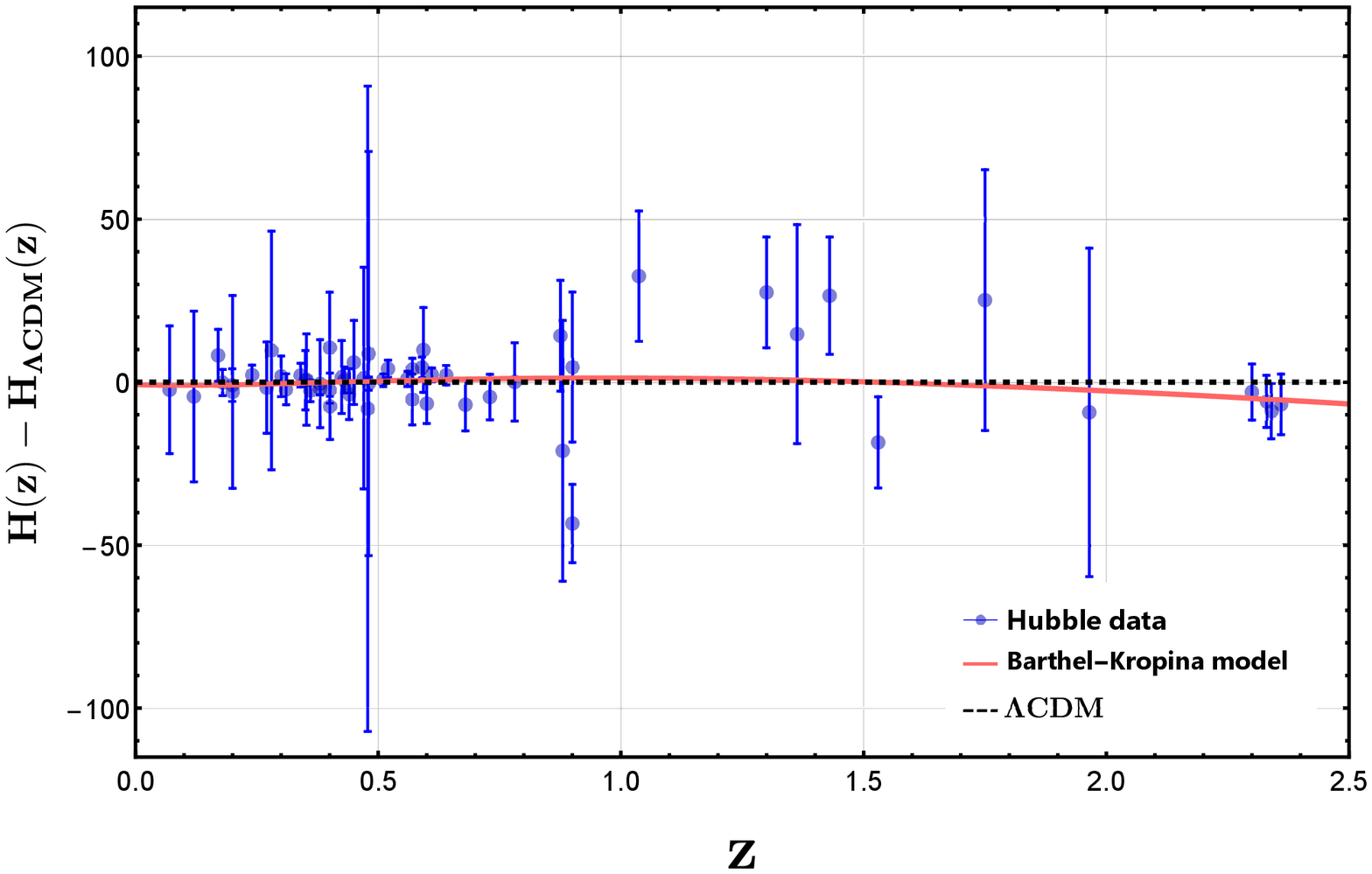}
\caption{The variation of the difference between the Barthel-Kropina dark energy model, and the $\Lambda$CDM model as a function of the redshift $z$ against the Hubble measurements.}\label{H_diffa}
\end{figure}

\paragraph{Relative difference between Barthel-Kropina and $\Lambda$CDM.} Finally, in Fig~\ref{H_diffa}, we plot the relative difference between the Barthel-Kropina dark energy model, and $\Lambda$CDM standard paradigm. The Figure shows that for $z<1.4$, the Barthel-Kropina dark energy model and the standard $\Lambda$CDM model behave almost similarly. However, some differences between the two models do appear for $z>1.4$, and these differences do increase with the redshift. Hence, in order to make the Barthel-Kropina model consistent with the observations, high redshift corrections may be necessary, which would involve, for example, a redshift dependent parameter of the equation of state of the dark energy. The consideration of the matter pressure, and of the radiation component, may also alleviate the differences between the two models at high redshifts. On the other hand, the precise determination of the Hubble function at high redshifts may provide a powerful test for discriminating between these two cosmological models.

\subsection{Cosmographic analysis}

Cosmographic analysis provides an universal and effective way to compare the solutions of the theoretical models with the cosmological observations. From the observational data we obtain a set of cosmological parameters, which must be compared with the predicted values of the same parameters, obtained from a given model. The result of the comparison allows us to conclude on the acceptability of the considered model. Thus, for a complete comparison of the Barthel-Kropina dark energy model with the observations, and the $\Lambda$CDM model, we will consider an extended set of parameters, constructed from the higher order time derivatives of the scale factor. More exactly, we will concentrate on the comparative behavior of the deceleration, jerk and snap parameters in the Barthel-Kropina and $\Lambda$CDM models.

\begin{figure}[htbp]
\centering
\includegraphics[scale=0.45]{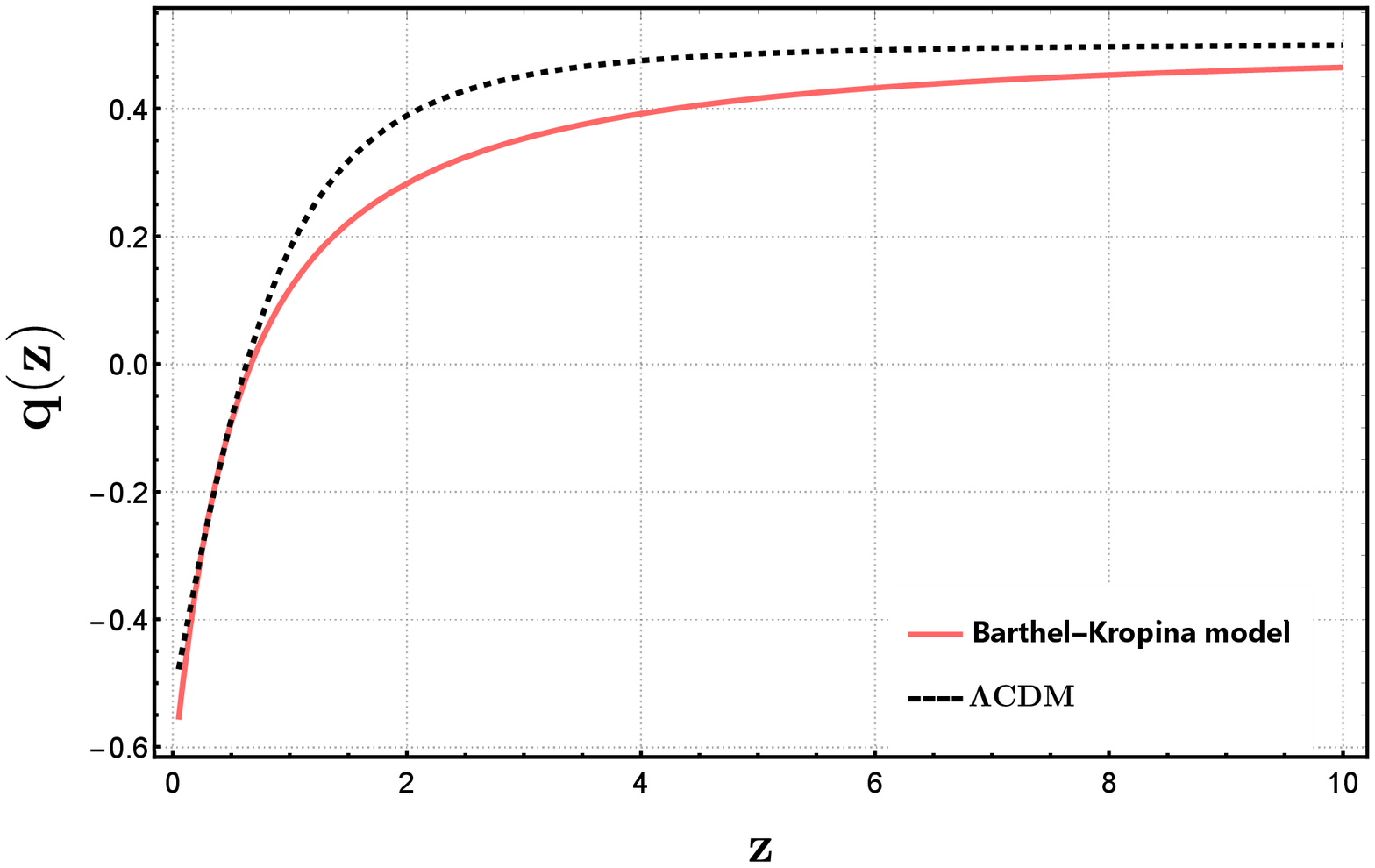}
\caption{Evolution of the deceleration parameter as a function of the redshift $z$ for the Barthel-Kropina and $\Lambda$CDM cosmologies.}\label{q_z}
\end{figure}

\paragraph{The deceleration parameter.} The redshift dependence of the deceleration parameter $q$ is represented comparatively for the Barthel-Kropina and for the $\Lambda$CDM models,  in Fig.~\ref{q_z}. The behavior of this parameter in the two models is almost identical in the redshift range $z\in (0,10)$, including the numerical value of the transition redshift $z_{tr}$ from the decelerating to the accelerating phase. However, the large time behavior of the two models is very different: while in the $\Lambda$CDM model the Universe ends in a de Sitter phase, with $q=-1$, in the Barthel-Kropina cosmology a super-accelerated evolution does occur, with $q(-1)\approx -2.7$.

\begin{figure}[htbp]
\centering
\includegraphics[scale=0.45]{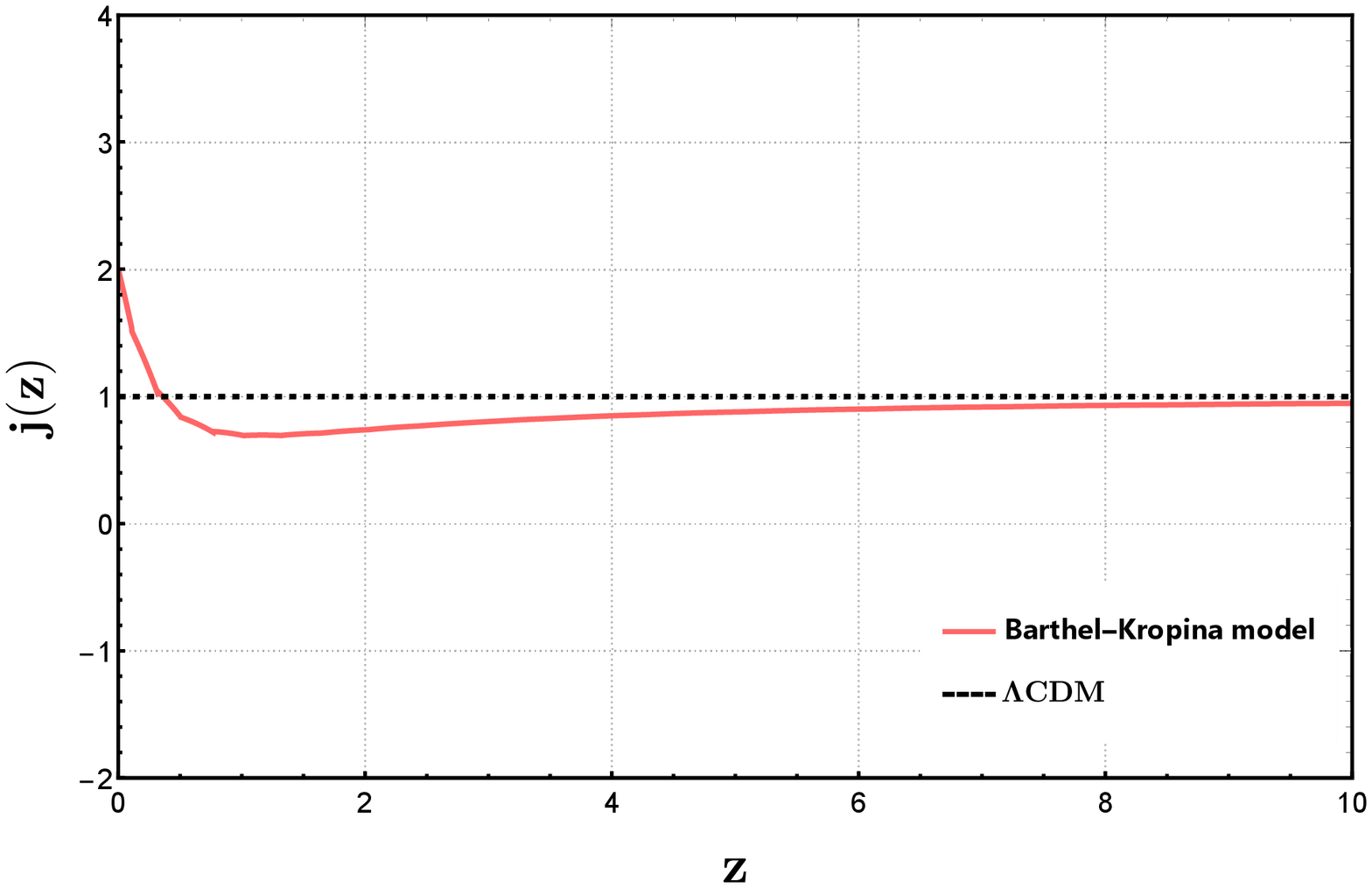}
\caption{Evolution  of the jerk parameter as a function of the redshift $z$ for the Barthel-Kropina and $\Lambda$CDM cosmological models.}\label{Jerk}
\end{figure}

\paragraph{The jerk parameter,} The redshift evolution of the jerk parameter $j(z)$ is represented in Fig.~\ref{Jerk}. At high redshifts $z>6$ the predictions of the two models basically coincide. However, important differences do exist at lower redshifts, with a very significant difference appearing at $z=1$, where the Barthel-Kropina model predicts a value twice as high as the $\Lambda$CDM value. The observational determination of the present day value of $j$ may thus provide an important test of the Barthel-Kropina cosmological model.

\begin{figure}[htbp]
\centering
\includegraphics[scale=0.42]{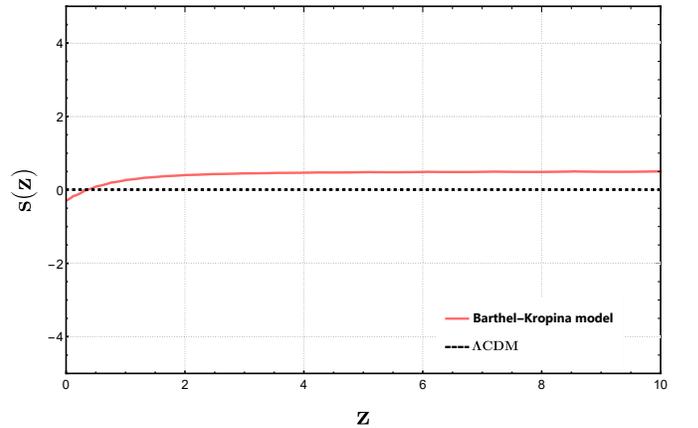}
\caption{The evolution of the snap parameter $s$ as a function of the redshift $z$ in the Barthel-Kropina and $\Lambda$CDM models.}\label{s}
\end{figure}

\paragraph{The snap parameter.} The variation with redshift of the snap parameter $s(z)$ is represented in Fig.~\ref{s}. In this case there is a systematic difference between the numerical values of $s$ in the Barthel-Kropina and the $\Lambda$CDM models, extending from low to high redshifts. Similarly to the $\Lambda$CDM model, $s$ is a constant in the Barthel-Kropina cosmology for $z>2$, but becomes a slightly increasing function for $z$ in the range $z\in (0,2)$, with the value of $s(0)$ having some differences in the two models.

\begin{figure}[htbp]
\includegraphics[scale=0.35]{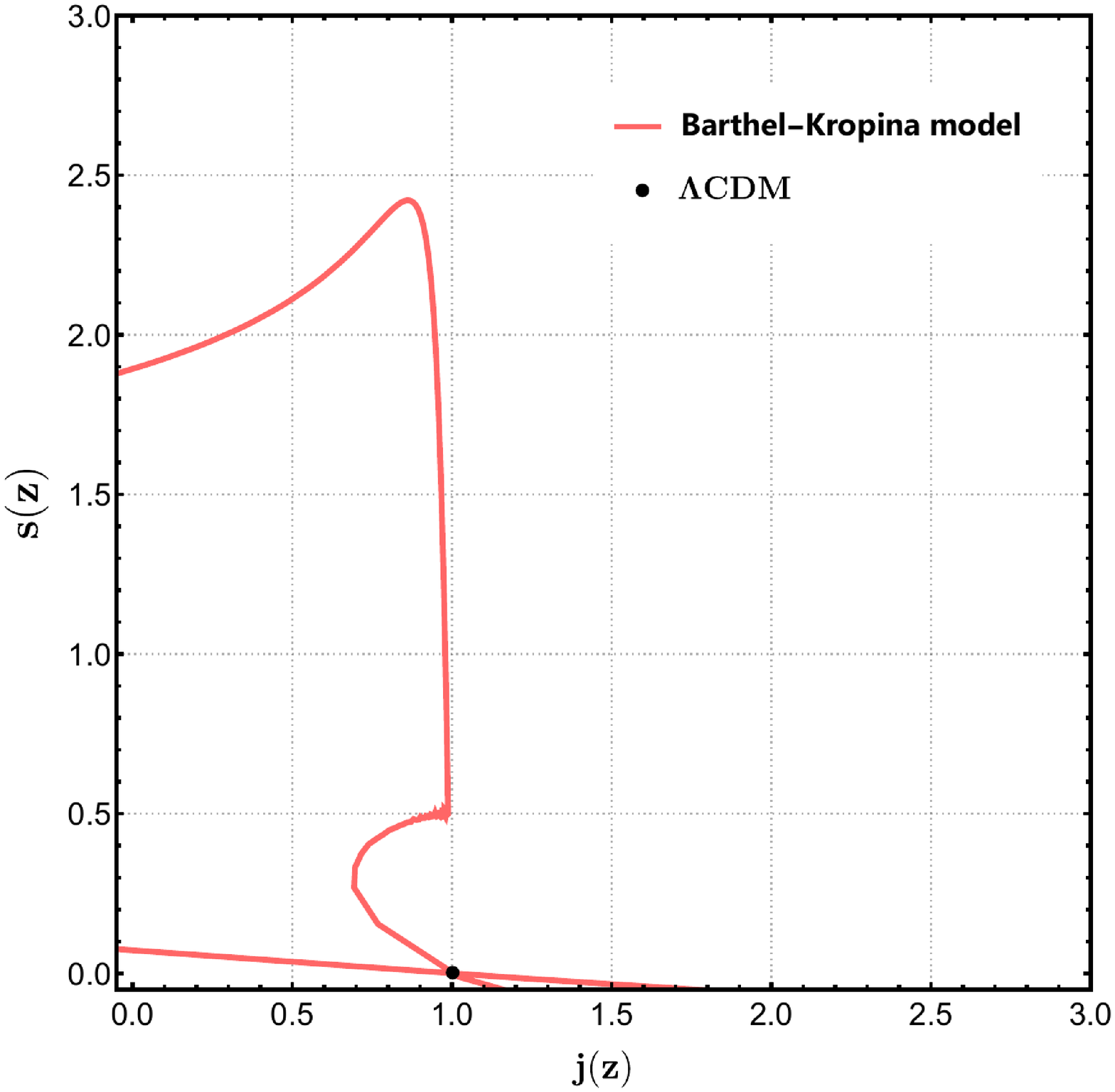}
\caption{The evolution of the snap parameter $s$ as a function of the jerk parameter $j$ in the Barthel-Kropina and $\Lambda$CDM models.}\label{js}
\end{figure}

\begin{figure}[htbp]
\centering
\includegraphics[scale=0.35]{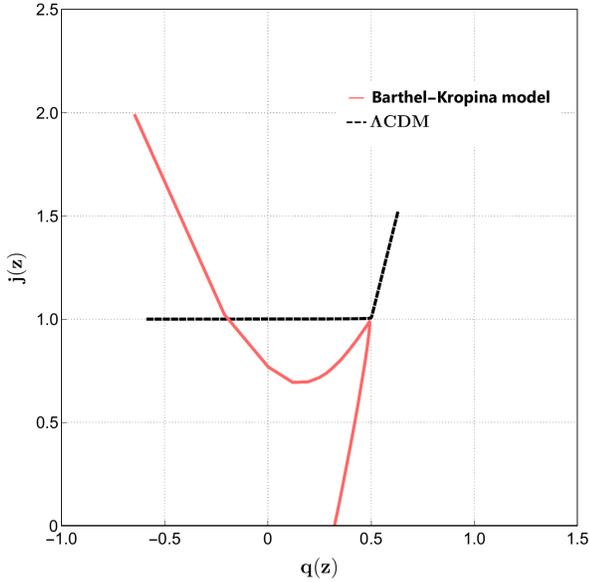}
\caption{The evolution of the jerk parameter $j$  as a function of the deceleration parameter $q$ in the Barthel-Kropina and $\Lambda$CDM models.}\label{qj}
\end{figure}

However, the differences between the predictions of the two models are highlighted more strongly when considering the parametric dependence of the snap parameter on the jerk parameter, and of the jerk parameter on the deceleration parameter, respectively. The functions $s=s(j)$ and $j=j(q)$ are represented in Figs.~\ref{js} and \ref{qj}, respectively, indicating a significant difference between the Barthel-Kropina and $\Lambda$CDM models. Thus, the observational determination of these cosmographic relations could offer a strong test of the validity of Finsler geometrical cosmological models.

\subsubsection{Cosmological quantities}

\begin{figure}[htbp]
\centering
\includegraphics[scale=0.47]{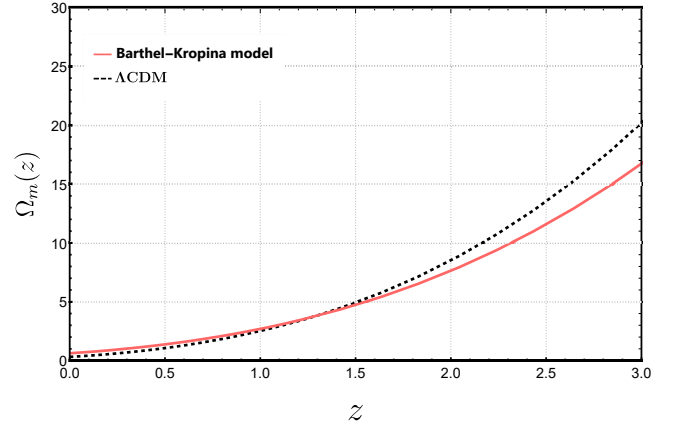}
\caption{The evolution of the matter density parameter as a function of the redshift $z$.}\label{Matter density}
\end{figure}

\begin{figure}[htbp]
\centering
\includegraphics[scale=0.42]{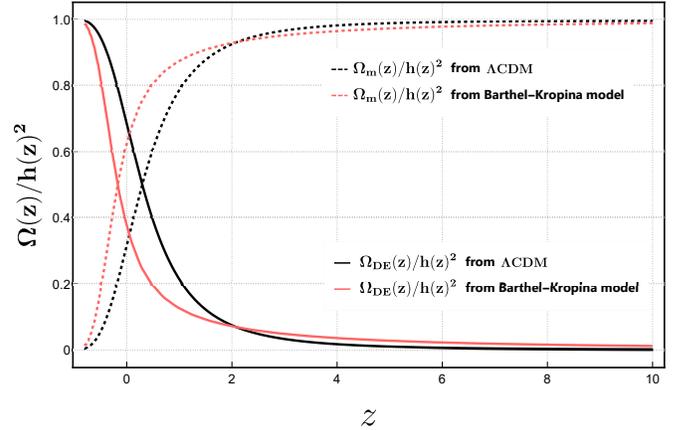}
\caption{The reduced matter density parameters  as a function of the redshift $z$ in the Barthel-Kropina and $\Lambda$CDM cosmological models.}\label{Reduced Om}
\end{figure}

\paragraph{Matter densities.} We consider now the comparative behavior of the relevant physical parameters of the Barthel-Kropina and $\Lambda$CDM models. The evolution of the matter density parameters $\Omega_m$ are presented in Fig.~\ref{Matter density}. Up to redshift of around $z\approx 2$ the matter densities in the two models almost coincide. However, at higher redshifts, significant differences begin to appear in the baryonic matter distributions, with the Barthel-Kropina model predicting a higher value of the matter density parameter. The comparisons of the reduced matter and dark energy density parameters in the two models are presented in Fig.~\ref{Reduced Om}. There is a good concordance between the theoretical predictions of both models for redshifts $z>2$. However, at low redshifts, significant differences in the behaviors of the reduced matter and dark energy density parameters can be seen, leading, at least in principle, to another possibility of observationally testing the Bearthel-Kropina dark energy model.

\paragraph{Dark energy properties.} When introducing the Barthel-Kropina dark energy model we have defined formally a geometric energy density, and a geometric pressure, related by a linear barotropic equation of state. However, in order to interpret physically these quantities, we need to take into account the energy conditions, that requires that the energy density must be positive. Hence, ion order to obtain from the geometric quantities physical quantities, we redefine them, in order to obtain a positive physical dark energy density, so that $\rho_{DEphy}=-\rho_{DE}$, and $p_{DEphy}=-p_{DE}$, respectively.

\begin{figure}[htbp]
\centering
\includegraphics[scale=0.4]{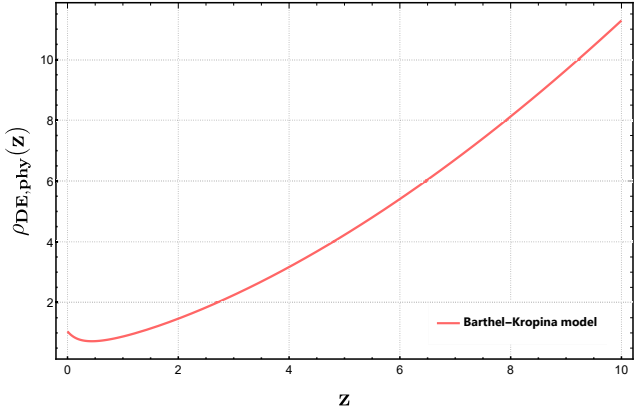}
\caption{The evolution of the physical energy density $\rho_{DEphy}$ in the Barthel-Kropina dark energy model, as a function of the redshift $z$.}\label{rho_DE}
\end{figure}

\begin{figure}[htbp]
\centering
\includegraphics[scale=0.40]{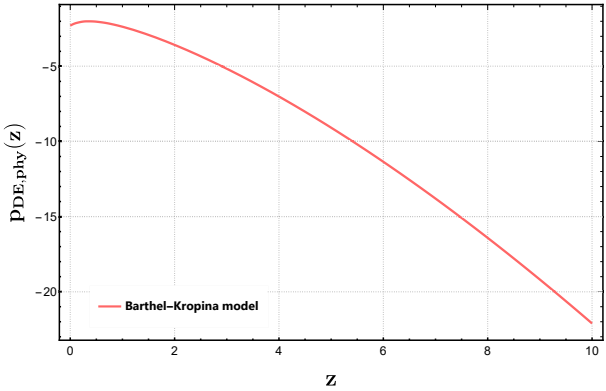}
\caption{The evolution of the physical pressure $p_{DEphy}$ in the Barthel-Kropina dark energy model as a function of the redshift $z$.}\label{pDE}
\end{figure}

The variation of the physical dark energy density is represented in Fig.~\ref{rho_DE}.  The physical Finsler type dark energy density is positive for all redshift values, and increases almost linearly for large redshift values. On the other hand, the effective physical pressure $p_{DEphy}$, depicted in Fig.~\ref{pDE}, is negative, and it is a monotonically decreasing function of the redshift. This situation is similar to most of the dark energy models, in which the energy density is positive, and the accelerating evolution of the Universe is triggered by a negative pressure. The properties of the physical dark energy density and pressure may have important implications for the further observational testing of the Finsler geometry based cosmological models.

\begin{figure}[htbp]
\centering
\includegraphics[scale=0.44]{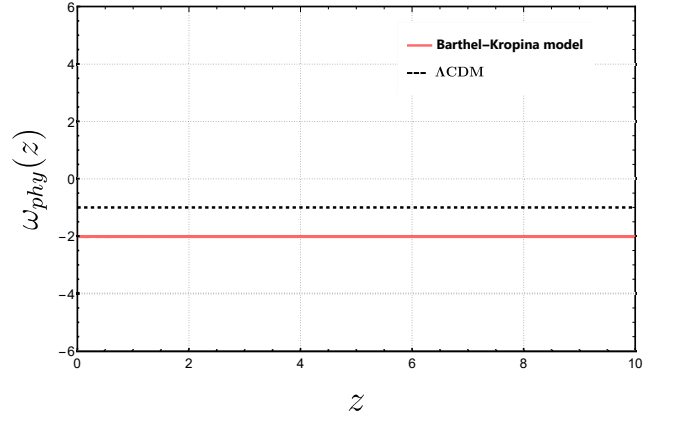}
\caption{The evolution of the equation of state parameter $\omega _{phys}$ as a function of the redshift $z$.}\label{EOS}
\end{figure}

The comparison between the dark energy equation of state parameters in the $\Lambda$CDM and the Barthel-Kropina models is represented in Fig.~\ref{EOS}. While in the $\Lambda$CDM model $\omega =-1$, in the Barthel-Kropina model $\omega =-2$, double the value of the standard cosmological model.

\paragraph{The coefficient of the one form $\beta$.} The variation of the coefficient $\beta$ of the Kropina cosmological metric is depicted in Fig.~\ref{Eta}. $\eta$ is a linearly  monotonically increasing function of $z$, which takes a finite value at the present time.
\begin{figure}[htbp]
\centering
\includegraphics[scale=0.49]{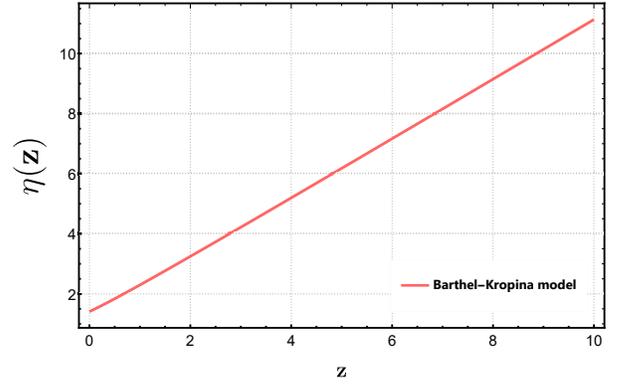}
\caption{The evolution of the coefficient $\eta(z)=(1+z)(1+\psi (z))$ of the one form $\beta$ of the Kropina metric.}\label{Eta}
\end{figure}

\paragraph{$Om(z)$ diagnostic}
The $Om(z)$ diagnostic \cite{Sahni:2008xx} is a powerful tool used to distinguish alternative cosmological models from the $\Lambda$CDM model. The $Om(z)$ function is defined as
\begin{equation}\label{om}
Om(z)=\frac{\left( H(z)/H_0\right)^{2}-1}{(1+z)^{3}-1}.
\end{equation}

For  the $\Lambda$CDM model, the function $Om(z)$ is always constant and equal to the present matter density $\Omega_{m0}$. Moreover, for models having a constant EoS, i.e. $\omega={\rm constant}$,  a positive slope of $Om(z)$ indicate a phantom behaviour,  while a negative slope corresponds  to a quintessence like behaviour. In the case of the standard general relativistic cosmology, and in the presence of a dark energy term obeying a linear barotropic equation of state, with the equation of state parameter denoted by $\omega$, the first Friedmann equation can be written as
\bea
\left(\frac{H(z)}{H 0}\right)^{2}&=&\Omega_{m 0}(1+z)^{3}+\left(1-\Omega_{m 0}\right) \nonumber\\
&&\times \operatorname{Exp}\left[3 \int_{0}^{z} \frac{1+\omega\left(z^{\prime}\right)}{1+z^{\prime}} d z^{\prime}\right].
\eea

For a constant $\omega$, we obtain
\be
\left(\frac{H(z)}{H 0}\right)^{2}=\Omega_{m 0}(1+z)^{3}+\left(1-\Omega_{m 0}\right)(1+z)^{3(1+\omega)} .
\ee
Hence, it follows that,
\be
Om(z)=\frac{\Omega_{m 0}(1+z)^{3}+\left(1-\Omega_{m 0}\right)(1+z)^{3(1+\omega)}-1}{(1+z)^{3}-1} .
\ee
For the $\Lambda \mathrm{CDM}$ model we have $\omega=-1$, and thus,
\be
Om(z)=\Omega_{m 0} .
\ee

By assuming that the Hubble function of the Barthel-Kropina model differs little from the $\Lambda$CDM one, so that $H^{BK}(z)=H^{\Lambda {\rm CDM}}(z)+\Delta H(z)$, with $\Delta H(z)/H^{\Lambda {\rm CDM}}(z)<<1$, from Eq.~(\ref{om}) we obtain, by neglecting the second order terms in $\Delta H$, the approximate expression
\bea
Om^{BK}(z)&\approx & Om^{\Lambda {\rm CDM}}(z)\nonumber\\
&&+\frac{2}{(1+z)^3-1}\frac{H^{\Lambda {\rm CDM}}(z)}{H_0^2}\Delta H(z)\nonumber\\
&=&Om^{\Lambda {\rm CDM}}(z)+\Delta Om^{BK}(z).
\eea

The variations with respect of the redshift of the $Om(z)$ functions on both Barthel-Kropina and $\Lambda$CDM cosmological models are represented in Fig.~\ref{Eta}.

\begin{figure}[htbp]
\centering
\includegraphics[scale=0.4]{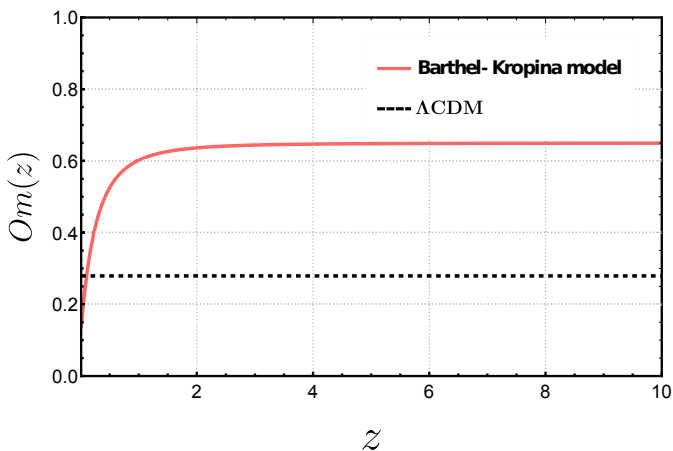}
\caption{The evolution of the $Om(z)$ function in the Barthel-Kropina and $\Lambda$CDM models.}\label{Eta}
\end{figure}

The $Om(z)$ functions are rather different for the two considered cosmological models. While for $\Lambda$CDM $Om(z)$ is an absolute constant, for redshifts in the range $z<2$, $Om(z)$ for the Barthel-Kropina model is an increasing function, reaching a constant value for $z>2$. In the constant region range, the numerical value of $Om(z)$ for the Barthel-Kropina cosmology is around two times higher than the corresponding $\Lambda$CDM value.

\section{Discussions and final remarks}\label{sect5}

In the present paper we have considered a detailed comparison of the theoretical predictions of a Finsler geometric type cosmological model, and the observations. More exactly, we have adopted as the basic metric function of our model the Kropina metric, which is a particular form of the general $(\alpha, \beta)$ metrics. Kropina spaces have many applications in physics, like, for example, quantum mechanics \cite{Tave3}, dissipative classical mechanics \cite{App1}, liquid crystal under
the influence of an external electromagnetic field \cite{App2}, nonequilibrium thermodynamic systems \cite{App3}, and in the study of nonlinear path of fluids flow through
inhomogeneous media \cite{App4}. However, the first investigations of the possible relevance of the Kropina geometry for the description of the gravitational phenomena were initiated in \cite{Fc26}, where the generalized Friedmann equations, describing the cosmological evolution, were obtained. In this Finslerian type approach it is assumed that the gravitational field is described by a Riemannian metric $g(x)$, which satisfies the Einstein gravitational field equations. However, the standard Einstein theory can be extended by nonlocalizing (anisotropizing) the background Riemann geometry, by attaching to each point $x =x^I$,  $I = 0, 1, 2, 3$,
an internal variable $y =y^J$, $J = 0, 1, 2, 3$. By further assuming that the internal variable $y$ is a vector, the nonlocalized Riemann geometry becomes a Finsler type geometry, leading to a Finsler type description of the gravitational interaction. Alternatively, one can describe the geometric properties of the Fisnlerian gravity by using the geometry of a general vector bundle.

In the nonlocal standard Finsler geometry the metric tensor $\hat{g}$ is a function of both local coordinates $x$ and of the internal vector $y$, so that $\hat{g} = \hat{g}(x, y)$. But in many realistic physical situations one can assume that the internal variable $y$ depends explicitly on the position, so that $y = Y (x)$. Therefore, the Finslerian metric becomes $\hat{g} = \hat{g} (x, Y (x)))$. Thus, in this type of geometric and physical models, the Finsler metric tensor $\hat{g}$ becomes a function
of $x$ only. The corresponding geometric structure is called the osculating Finsler manifold. In this particular osculating Finsler space one can introduce a specific connection, which is called the  Barthel connection, which has the remarkable property that in the case of the Kropina geometry, and in all $(\alpha, \beta)$ type geometries) it is the Levi-Civita connection of the Riemannian metric $\hat{g}(x)=\hat{g}\left(x, Y(x)\right))$.

By adopting for the background Riemann metric the Friedmann-Lemaitre-Robertson-Walker (FLRW) form, one can obtain the generalized Friedmann equations of Barthel-Kropina cosmological model. These equations are mathematically simple, and, after an appropriate choice of the coefficient of the one-form $\beta$, they closely resemble the Friedmann equations of standard general relativity, but also contain extra terms generated by the Finsler geometric effects. In the present approach, for simplicity, we interpret these geometric terms as describing dark energy only, even that the possibility they describe both dark matter and dark energy could be also considered. In order to close the system of cosmological equations we have imposed a linear relation between the geometric dark energy and dark pressure terms. Hence, with this choice the system of generalized field equations can be closed, and it can be formulated as a dynamical system in the redshift space. However, the solution of the system can be obtained only numerically, and it depends on three model parameters: the present day values of one form component $\eta$ and of its derivative, and of the parameter of the geometric equation of state $\omega$.

In the present work we have performed a systematic and rigorous analysis of the cosmological implications of the Barthel-Kropina dark energy model, by comparing it with the cosmological observations for 57 Hubble function values, and for the Pantheon dataset. To obtain the best fit values of the model parameters we have used the MCMC method, from which one can infer the optimal fits for the model parameters. The parameter of the geometric dark energy equation of state turns out to be of the order $\omega \approx 2.5$, while the present day values of $\eta$ are obtained as $\eta (0)\approx 1.45$, and $\eta '(0)\approx 1.65$, respectively. Hence, our results indicate the presence of significant Finsler geometric effects in the $z=0$ Universe. By using the optimal fits values, the  fittings of the Hubble function data give very good results, for both the Hubble and the Pantheon data. Moreover, there is a very close relationship between the Barthel-Kropina and the $\Lambda$CDM cosmologies, with both overlapping for small ($z<1.5$) redshift values. The comparison of the two models has also been performed in a more quantitative way by studying the Akaike Information Criterion, which indicates that the Barthel-Kropina cosmology is rather closed to the standard general relativistic description in the presence of a cosmological constant.

In order to obtain a better estimate of the strengths and weaknesses of the Barthel-Kropina cosmology we have also performed a detailed comparison of the cosmographic parameters, and of other relevant cosmological quantities. If the behavior of the deceleration parameter agrees well with that of the standard $\Lambda$CDM, significant differences do appear between models when one considers the jerk and snap parameters. Hence, the cosmographic approach can offer the possibility of discriminating between Riemann and Finsler geometry based cosmological models. The variations of the matter and density parameters of the two models are also relatively similar, and even coincide on some redshift ranges.

There are several possibilities for explaining the recent cosmological observational data,  which can be described as the dark components approach, the dark geometry approach, and the dark couplings approach, respectively \cite{LoHa}. The present investigation of the evolution and dynamics of the Universe, including the study of the accelerating expansion, has been done in the framework of the dark geometry approach, by assuming that the true geometry of the Universe goes far beyond the Riemann geometry of general relativity, and that the extra geometric terms coming from the post-Riemannian mathematical structures may be responsible for the presence of dark matter and dark energy in the Universe. In the present work we have investigated in detail, from the point of view of the consistency with observations, an example
of a dark geometric model, the osculating Barthel-Kropina-FLRW geometry, which has its roots in the Finsler geometry. In this model,  an effective fluid type dark energy can be generated from the geometric structures underlying the dark Barthel-Kropina geometry. An interesting property of this model is the close relation between the general relativistic Friedmann cosmological evolution equations, and the Barthel-Kropina ones, with this relation allowing to introduce in a natural way a fluid type geometric dark
energy term for the description of the gravitational dynamics. The model also gives a very good description of the observational data in terms of only three free parameters, the present day values of the coefficients of the one form $\beta$, and the parameter of the equation of state. It also almost exactly reproduces the $\Lambda$CDM model predictions, but significant differences still exist for high redshifts, and in the values of some cosmographic parameters.
Therefore, the Barthel-Kropina-FLRW cosmological model could become an appealing geometric alternative to the standard $\Lambda$CDM model in terms of the explanations of the observational data. It could also provide some new insights, and a better understanding of the complex interaction between abstract mathematical structures,  and the physical reality.

\section*{Acknowledgments}

We would like to thank the anonymous Referee for comments and suggestions that helped us to improve our work. The work of TH is supported by a grant of the Romanian Ministry of Education and Research, CNCS-UEFISCDI, project number PN-III-P4-ID-PCE-2020-2255 (PNCDI III).

\end{document}